\documentclass[twocolumn,superscriptaddAppendicesmsmath,amstex,amssymb,citeautoscript,nofootinbib,superscriptaddress]{revtex4-1}

\usepackage[english]{babel}
\usepackage{letltxmacro}
\usepackage{latexsym}
\usepackage[utf8]{inputenc}
\LetLtxMacro{\ORIGselectlanguage}{\selectlanguage}
\makeatletter
\DeclareRobustCommand{\selectlanguage}[1]{%
  \@ifundefined{alias@\string#1}
    {\ORIGselectlanguage{#1}}
    {\begingroup\edef\x{\endgroup
       \noexpand\ORIGselectlanguage{\@nameuse{alias@#1}}}\x}%
}
\newcommand{\definelanguagealias}[2]{%
  \@namedef{alias@#1}{#2}%
}
\makeatother
\definelanguagealias{en}{english}
\definelanguagealias{English}{english}
\usepackage{graphicx}
\usepackage{amsmath}
\usepackage{amsfonts}
\usepackage{amsthm}   
\usepackage{amssymb}
\usepackage{bm}
\usepackage{color}
\usepackage[percent]{overpic}
\usepackage{soul} 
\usepackage{amssymb}
\usepackage{wasysym}
\usepackage{dsfont}
\usepackage{float}
\usepackage{mathtools}
\usepackage{enumitem}
\usepackage{dsfont}
\usepackage{multirow}
\usepackage{colortbl}
\usepackage{mathrsfs}
\usepackage{makecell}
\usepackage{physics}
\makeatletter
\let\old@makecaption=\@makecaption
\usepackage{subcaption}
\let\@makecaption=\old@makecaption
\makeatother

\usepackage{hyperref}
\hypersetup{
    bookmarks=false,         
    unicode=false,          
    pdftoolbar=false,        
    pdfmenubar=true,        
    pdffitwindow=false,     
    pdfstartview={FitH},    
    pdftitle={},    
    pdfauthor={Authors},     
    pdfsubject={},   
    pdfcreator={},   
    pdfproducer={}, 
    pdfkeywords={many-body localization} {matrix elements} {disordered systems}, 
    pdfnewwindow=true,      
    colorlinks=true,       
    linkcolor=blue,          
    citecolor=blue,        
    filecolor=magenta,      
    urlcolor=black           
}

\theoremstyle{definition}

\newcommand{\be}{\begin{equation}}
\newcommand{\ee}{\end{equation}}
\newcommand{\bea}{\begin{eqnarray}}
\newcommand{\eea}{\end{eqnarray}}

\renewcommand{\bar}[1]{\overline{ #1}}

\newcommand{\ri}{\mathrm{i}}
\newcommand{\rd}{\mathrm{d}}

\newcommand{\Langle}{\langle\!\langle}
\newcommand{\Rangle}{\rangle\!\rangle}

\usepackage{graphicx}
\usepackage[colorinlistoftodos]{todonotes}
\usepackage{verbatim}

\newcommand{\bbZ}{\mathbb{Z}}

\newcommand{\bbG}{\mathbb{G}}

\newcommand{\calA}{\mathcal{A}}

\newcommand{\calC}{\mathcal{C}}

\newcommand{\calG}{\mathcal{G}}
\newcommand{\calH}{\mathcal{H}}

\newcommand{\calL}{\mathcal{L}}
\newcommand{\calM}{\mathcal{M}}
\newcommand{\calN}{\mathcal{N}}
\newcommand{\calO}{\mathcal{O}}
\newcommand{\calP}{\mathcal{P}}

\newcommand{\calR}{\mathcal{R}}
\newcommand{\calS}{\mathcal{S}}

\newcommand{\calW}{\mathcal{W}}

\newcommand{\calZ}{\mathcal{Z}}

\newcommand{\sfa}{\mathsf{a}}
\newcommand{\sfb}{\mathsf{b}}
\newcommand{\sfc}{\mathsf{c}}
\newcommand{\sfp}{\mathsf{p}}

\newcommand{\sfg}{\mathsf{g}}

\newcommand{\bfM}{\mathbf{M}}

\newcommand{\bfr}{\mathbf{r}}

\usepackage{tikz}
\usepackage{tikz-cd}
\usetikzlibrary{arrows}
\usetikzlibrary{intersections}
\usetikzlibrary{shapes.geometric}
\usetikzlibrary{decorations.pathmorphing, patterns,shapes}
\usetikzlibrary{decorations.markings}
\usetikzlibrary{patterns}

\tikzset{
	mid arrow/.style={postaction={decorate,decoration={
				markings,
				mark=at position .575 with {\arrow[#1]{stealth}}
	}}},
	near arrow/.style={postaction={decorate,decoration={
				markings,
				mark=at position .275 with {\arrow[#1]{stealth}}
	}}},
	far arrow/.style={postaction={decorate,decoration={
				markings,
				mark=at position .800 with {\arrow[#1]{stealth}}
	}}},
}

\usepackage{soul}

\newtheorem*{lemma*}{Lemma}

\begin{document}

\title{Information dynamics in decohered quantum memory with repeated syndrome measurements: a dual approach}

\author{Jacob Hauser}
\affiliation{Department of Physics, University of California, Santa Barbara, CA 93106, USA}

\author{Yimu Bao}
\affiliation{Kavli Institute for Theoretical Physics, University of California, Santa Barbara, CA 93106, USA}

\author{Shengqi Sang}
\affiliation{Kavli Institute for Theoretical Physics, University of California, Santa Barbara, CA 93106, USA}
\affiliation{Perimeter Institute for Theoretical Physics, Waterloo, Ontario N2L 2Y5, Canada}
\affiliation{Department of Physics and Astronomy, University of Waterloo, Waterloo, Ontario N2L 3G1, Canada}

\author{Ali Lavasani}
\affiliation{Kavli Institute for Theoretical Physics, University of California, Santa Barbara, CA 93106, USA}

\author{Utkarsh Agrawal}
\affiliation{Kavli Institute for Theoretical Physics, University of California, Santa Barbara, CA 93106, USA}

\author{Matthew P. A. Fisher}
\affiliation{Department of Physics, University of California, Santa Barbara, CA 93106, USA}

\begin{abstract}
Measurements can detect errors in a decohered quantum memory allowing active error correction to increase the memory time.
Previous understanding of this mechanism has focused on evaluating the performance of error correction algorithms based on measurement results.
In this work, we instead intrinsically characterize the information dynamics in a quantum memory under repeated measurements, using coherent information and relative entropy.
We consider the dynamics of a $d$-dimensional stabilizer code subject to Pauli errors and noisy stabilizer measurements and develop a $(d+1)$-dimensional statistical mechanics model for the information-theoretic diagnostics. 
Our model is dual to the model previously obtained for the optimal decoding algorithm, and the potential decoding transition in the quantum memory again manifests as a thermal phase transition in the statistical mechanics model.
We explicitly derive the model and study the phase transition in information encoding in three examples: surface codes, repetition codes, and the XZZX code.

\end{abstract}

\maketitle

\section{Introduction}\label{sec:intro}
Protecting quantum information against decoherence in realistic devices is a central focus of quantum information science.
While designing a passive memory that encodes quantum information in thermal equilibrium states remains elusive below four dimensions, one can alternatively perform active error correction to protect encoded information in non-equilibrium states~\cite{brown2016quantum,terhal2015quantum}.
This entails using repeated measurements to detect error syndromes and applying feedback based on the measurement results to correct errors.
Measurements remove entropy from the system and therefore can compete with decoherence to increase the memory time, even if the measurement outcomes are noisy.

Prior work has sought to understand this mechanism by examining the performance of specific decoding algorithms for specific quantum error-correcting (QEC) codes.
For example, in a surface code subject to decoherence, repeated noisy syndrome measurements enable a decoding algorithm to protect quantum information until exponentially late times as long as the rates of decoherence and measurement noise are below a finite threshold~\cite{dennis2002topological}.
This is demonstrated in Ref.~\cite{dennis2002topological} by mapping the transition in the fidelity of the decoding algorithm to the confinement transition in the 3D random-plaquette Ising model (RPIM), i.e.\ a random $\bbZ_2$ gauge theory.
It is natural to ask whether the impact of repeated syndrome measurements on memory time can be understood intrinsically from the information dynamics of the system without reference to a particular decoding algorithm.

It has been recently shown that the surface code with local decoherence can undergo an intrinsic information transition in the mixed density matrix of the system, which governs the decoding transition based on a single round of perfect measurements~\cite{fan2023diagnostics,bao2023mixed,lee2023quantum,wang2023intrinsic,su2024tapestry,lyons2024understanding,zhao2023extracting,colmenarez2023accurate,lee2024exact,sohal2024noisy}.
In particular, Ref.~\cite{fan2023diagnostics} maps the coherent information (the amount of recoverable information from the mixed state), along with several other information diagnostics, to observables in a 2D multi-flavor Ising model.
At a finite decoherence rate, the multi-flavor Ising model undergoes a ferromagnetic transition, detected by the information diagnostics.
Notably, the multi-flavor Ising model is dual to the random-bond Ising model along the Nishimori line~\cite{nishimori1981internal}, which characterizes the decoding transition for the optimal decoding algorithm~\cite{dennis2002topological}. 
This duality stems from two different expansions of the same decohered density matrix~\cite{fan2023diagnostics}.

In this work, we extend this perspective to general stabilizer codes subject to local decoherence and repeated imperfect syndrome measurements.
We study information dynamics characterized by intrinsic information diagnostics, including coherent information and relative entropy.
To this end, we map the R\'enyi versions of the diagnostics to observables in statistical mechanics (stat-mech) models based on the stabilizer expansion of the physical density matrix.
We further perform the expansion of the same density matrix in terms of error configurations and obtain the stat-mech model previously derived for the optimal decoding algorithm (i.e.\ the maximum likelihood decoder)~\cite{dennis2002topological,Wang:2002ph,Katzgraber:2009zz,Bombin:2012jk,Kubica:2018rab,Chubb:2021htd,Song:2021bud,Venn:2022kxy,Behrends:2022feh}.
The two stat-mech models are related by high-to-low temperature duality.
In the example of decohered surface codes under repeated syndrome measurements, the decoding transition manifests as a ferromagnetic transition in a 3D multi-flavor Ising model (in the stabilizer expansion) dual to the confinement transition in the 3D RPIM (in the error configuration expansion).

\subsection{Overview}
\begin{figure}[t!]
    \centering
    \begin{tikzpicture}[]
    \pgftext{
    \includegraphics[width=0.5\linewidth]{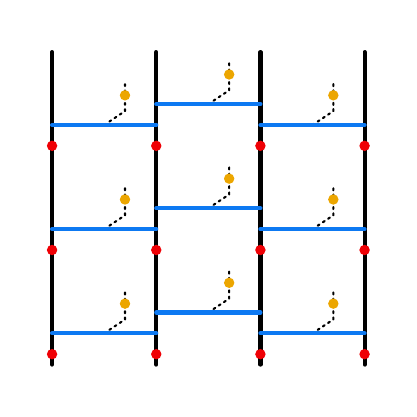}%
    \includegraphics[width=0.5\linewidth]{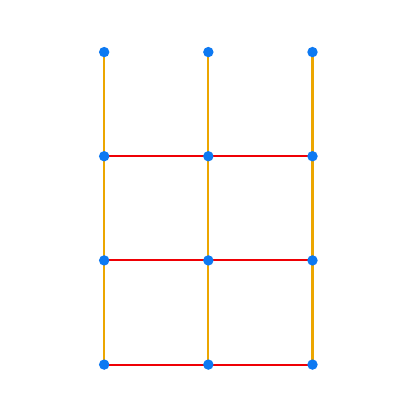}
    } at (0pt,0pt);
    \draw[->,>=stealth] (-4.2,-1) -- (-4.2,1) node[above]{$t$};
    \node at (-2.15, -2.1) {(a)};
    \node at (2.15, -2.1) {(b)};
    \end{tikzpicture}
    \caption{(a)~One-dimensional repetition code under repeated measurements. Each black solid line denotes a physical qubit. In every time step, check operators $Z_rZ_{r+1}$ on neighboring qubits are measured (horizontal blue lines) with measurements modeled by coupling measurement ancillae (black dashed lines) to physical qubits. The red and orange circles denote the physical decoherence and measurement errors on the physical qubits and ancillae, respectively.
    (b)~The 2D multi-flavor Ising model for the 1D repetition code. The spins live on the vertices (blue dots) of a 2D square lattice; each vertex corresponds to a check operator in the circuit. The spatial couplings (in red) are related to the strength of physical decoherence. The temporal couplings (in orange) are related to the error rate of measurement readout.}
    \label{fig:overview_figure}
\end{figure}

We begin with an overview of our results before delving into the details.
We consider the dynamics of $d$-dimensional stabilizer codes subject to local Pauli errors and repeated stabilizer measurements. 
The dynamics is described by a quantum circuit of $T$ steps as illustrated in Fig.~\ref{fig:overview_figure}(a).
In each time step, we first apply single-qubit Pauli channels to the physical qubits and then perform noisy measurements on a set of stabilizers (called check operators).
To characterize the information dynamics, we consider the coherent information, the amount of recoverable information in the system together with the measurement record, and the relative entropy between the quantum states of the memory for two initial states.
The key result of this work is to develop a $(d+1)$-dimensional stat-mech model for the $n$-th moment $\tr \rho_{Q\bfM}^n$ of the density matrix of the system $Q$ together with measurement record $\bfM$ as illustrated in Fig.~\ref{fig:overview_figure}(b).
The mapping allows evaluating the R\'enyi versions of our information diagnostics as observables in the stat-mech model.

In Sec.~\ref{sec:stat-mech_stabilizer}, we detail the mapping from the $n$-th moment $\tr\rho_{Q\bfM}^n$ to the partition function of the stat-mech model based on the stabilizer expansion of $\rho_{Q\bfM}$.
The model involves $n-1$ flavors of Ising spins and has a Hamiltonian
\begin{equation}\label{eq:overview_full_hamiltonian}
    \calH^{(n)}(\{\bm{\sigma}^\sfa\}) = \sum_{\sfa} \calH(\bm{\sigma}^\sfa) + \calH(\textstyle\prod_\sfa \bm{\sigma}^\sfa),
\end{equation}
where $\bm{\sigma}^\sfa$ denotes a collection of the $\sfa$-th flavor of Ising spins.
The Hamiltonian $\calH(\bm{\sigma}^\sfa)$ for each flavor involves $T+1$ layers of Ising spins for $T$ time steps. 
Each spin $\sigma_{t,i}$ in the $t$-th layer is associated with a check operator $g_i$.
The Hamiltonian $\calH(\bm{\sigma})$ contains intralayer coupling $H(\bm{\sigma}_t)$ determined by the decoherence rate and interlayer Ising couplings with strength $J_q = -(1/2)\log(1-2q)$ controlled by the measurement error rate $q$, i.e.\
\begin{equation}
    \calH(\bm{\sigma}) = \sum_{t = 0}^{T-1} H(\bm{\sigma}_t) - J_q\sum_{t=0}^{T-1}\sum_i\sigma_{t,i}\sigma_{t+1,i},
\end{equation}
where we use $\bm{\sigma}_t$ and $\bm{\sigma} := \{\bm{\sigma}_t\}$ to denote the spin configuration at time $t$ and the entire spacetime, respectively.
We note that in this model, higher noise rates give rise to larger couplings (equiv. lower temperatures).

The multi-flavor Ising model exhibits a symmetry that depends on the redundancy in the set of measured check operators.
A subset of check operators is redundant if its product is the identity, and such redundancies form a group $\calR$.
The symmetry of the multi-flavor Ising model is then given by $\bbG^{(n)} = (\calR^n \rtimes S_n)/\calR$.
Notably, $\bbG^{(n)}$ is either a local or global symmetry depending on whether the redundant subset of check operators has local or global support. 
For example, in the 2D toric code under repeated measurements of plaquette and star operators, the redundancy $\calR = \bbZ_2 \times \bbZ_2$ is given by the product of all plaquette operators and the product of all star operators, resulting in a global symmetry of the stat-mech model.
In the 2D repetition code on a square lattice with check operators on the edges, the redundancy is given by the product of four check operators locally, giving rise to a local (gauge) symmetry of the stat-mech model.
When tuning the decoherence rate, the stat-mech model can undergo a phase transition that either spontaneously breaks the global symmetry $\bbG^{(n)}$ or confines the gauge field.

The information diagnostics map to distinct observables in the stat-mech model.
Specifically, the coherent information is associated with the excess free energy of inserting defects in the stat-mech model, and the relative entropy maps to correlation functions.
In the examples studied in this work (surface codes, repetition codes, and the XZZX code), such observables detect the underlying phase transition in the stat-mech model and exhibit singularities at the same point.

We note that the boundary condition in the stat-mech model depends on the diagnostic we consider.
For the information diagnostics associated with both the system $Q$ and the measurement record $\bfM$, the stat-mech model has an open boundary condition.
In contrast, the stat-mech model for the diagnostics associated with only the measurement record has an external field at the final layer of the spins.

The multi-flavor model obtained in the stabilizer expansion is dual to the stat-mech model previously derived for the maximum likelihood (ML) decoder.
In Sec.~\ref{sec:stat-mech_error_configuration}, we explicitly map the $n$-th moment $\tr\rho^n_{Q\bfM}$ to a stat-mech model based on the error configuration expansion.
Such a model reduces to that for the ML decoder in the limit $n \to 1$ and is the Kramers-Wannier dual of the multi-flavor Ising model.

Before proceeding, we comment on the advantages of the multi-flavor Ising model obtained in the stabilizer expansion.
First, the mapping excels at identifying codes with maximum threshold. 
In our model, the high-temperature phase corresponds to the low error rate regime of the stabilizer code.
Thus, a code with a maximum error threshold (e.g. $p_c = 0.5$ in the toric code with Pauli-Y error) corresponds to a stat-mech model with no finite-temperature ordering.
These models are better understood compared to their dual (obtained previously for the ML decoder), which are highly constrained and order up to infinite temperature.
Second, the impact of measurement error is clear in the multi-flavor Ising model.
A higher rate of measurement readout errors corresponds to a stronger interlayer coupling, which favors the ferromagnetic phase that does not encode information.
Third, the benefit of repeating noisy syndrome measurements in quantum error correction is also explicit in our mapping.
Successful error correction relies on using the measurement record to identify the error configurations in the system. 
The corresponding stat-mech model has an external field on the final layer of spins due to the measurement readout errors.
When considering QEC in the decohered 2D surface code, the presence of this symmetry-breaking field indicates the decodable phase (i.e.\ a paramagnetic phase) can exist only when the measurements are repeated an extensive number of times.

The rest of the paper is organized as follows.
Sec.~\ref{sec:setup} describes the dynamics of stabilizer code under repeated noisy syndrome measurements.
Sec.~\ref{sec:diagnostics} introduces the diagnostics of information dynamics in quantum memory.
Sec.~\ref{sec:stat-mech} develops stat-mech models for information diagnostics based on two expansions of the density matrix. 
We close with discussion in Sec.~\ref{sec:discussion}.

\tableofcontents

\section{Setup}\label{sec:setup}
In this work, we study the dynamics of quantum information in stabilizer codes under decoherence and repeated noisy syndrome measurements.
We keep track of the mixed density matrix describing the quantum state of the system together with the classical measurement outcomes. 
In what follows, we first review the definition of stabilizer codes and then detail the dynamics considered in this work.

Stabilizer codes are a broad class of QEC codes whose code space $\calL$ is the simultaneous $+1$ eigenspace of a set of mutually commuting Pauli operators $\sfg_i$~\cite{gottesman1997stabilizer}
\begin{equation}
    \calL = \mathrm{span}\{\dyad{\psi}{\psi} :
    \sfg_i|\psi\rangle=\ket{\psi},\, i \in 1, 2, \dots, I_0\}.
\end{equation}
The Pauli operators generate a stabilizer group under multiplication\footnote{Without loss of generality, we set the phase of each Pauli operator to be $+1$.}
\begin{align}
    \calS = \langle \sfg_1, \sfg_2, \dots, \sfg_{I_0}\rangle.
\end{align}
Each element $s \in \calS$ is called a \emph{stabilizer} and takes a $+1$ eigenvalue in the code space.
We take the $I_0$ generators to be independent; the stabilizer group then contains $2^{I_0}$ elements and is isomorphic to $\bbZ_2^{I_0}$.

A stabilizer code with $N$ physical qubits can encode $K = N - I_0$ logical qubits. 
We use $\bar{g}_k^x$, $\bar{g}_k^z$ for $k = 1, 2, \dots, K$ to denote the Pauli-X and -Z operators of logical qubits. 
These operators commute with the stabilizers and operate within the code space. 
In practice, stabilizer measurements are performed to identify errors in a stabilizer code; $s|\psi\rangle = -\ket{\psi}$ for any $s \in \calS$ indicates a departure from the code space. 
We refer to the collection of stabilizer measurement outcomes as an \emph{error syndrome}.

We are now ready to introduce the stabilizer code dynamics studied in this work.
We consider a stabilizer code with $N$ physical qubits initialized in an encoded state $\rho_0 \in \calL$ and evolved under decoherent interaction with the environment interrupted by repeated syndrome measurements (as shown in Fig.~\ref{fig:overview_figure}) for $T$ time steps. 
In particular, at each time step $t \in \{1,\dots,T\}$ the following occurs: 
\begin{enumerate}
    \item We apply single-qubit Pauli channels to every qubit in the system $Q$, with $p_x$, $p_y$, and $p_z$ denoting the probability of each type of Pauli error\footnote{Although we focus on uncorrelated Pauli errors in this work, our framework can be straightforwardly extended to correlated Pauli errors.}.
    \item We measure a set of stabilizers $\calC$ and record their measurement results in the ancilla qubits in $M_t$. Each measurement may have a readout error; its result is recorded incorrectly with probability $q$.
\end{enumerate}

The decoherence at each time step is described by $\calN[\rho] = \prod_r \calN_r[\rho]$ where the single-qubit Pauli channel takes the form
\begin{equation}\label{eq:noise_channel_def}
    \calN_r[\rho] = \left(1 - \sum_\alpha p_\alpha\right) \rho + \sum_\alpha p_\alpha P^\alpha_r \rho P^\alpha_r.
\end{equation}
Here, $P^{\alpha}_{r} = X_r, Y_r, Z_r$ for $\alpha = x,y,z$, and their corresponding error rates satisfy $p_\alpha > 0$ and $\sum_\alpha p_\alpha \leq 1$.
In this work, we focus on the regime $p_{\alpha_1} + p_{\alpha_2} \leq 1/2$ for any $\alpha_1 \neq \alpha_2$.

We perform measurements on the same set of check operators in the stabilizer group at each time step,
\begin{align}
    \calC := \{g_1, g_2, \dots, g_I\},
\end{align}
where each $g_i \in \calS$.
We focus on the case that $\calC$ generates the entire stabilizer group. 
However, the check operators in $\calC$ may form an overcomplete generating set with redundant check operators given by a product of the others. 
The redundancies in check operators lead to constraints in measurement results in the absence of measurement errors and may provide additional information when performing error correction with noisy measurement results.
These redundancies form a group $\calR$ that is isomorphic to $\bbZ_2^{I-I_0}$. 
In the 1D repetition code with a periodic boundary condition, the product of all check operators is the identity, giving rise to a global $\mathbb{Z}_2$ redundancy.
In the 2D repetition code on the square lattice, the product of check operators around a plaquette is the identity, giving rise to local $\mathbb{Z}_2$ redundancies.

A measurement can be modeled by coupling the system $Q$ to an ancilla $M$ via a unitary operator and then projectively measuring the ancilla. Specifically, the unitary operator acting on the system and ancilla prepared in $\ket{0}_M$ defines an isometric embedding
\begin{align}
    U_{Q \to QM} = \calP_0 \otimes \ket{0}_M + \calP_1 \otimes \ket{1}_M, \label{eq: U_Q,QM}
\end{align}
where $\calP_{0,1} = (1 \pm g)/2$ is the projector associated with the outcome $\pm 1$ when measuring a check operator $g$, and $(U_{Q\to QM})^\dagger U_{Q\to QM} = \mathds{1}_Q$. 
A subsequent measurement on ancilla $M$ in the Pauli-Z basis reads out the outcome and projects the system $Q$ to the corresponding eigenstate of the check operator $g$.
This measurement outcome is probabilistic with its probability determined by the Born rule, $p_{0,1} = \tr \calP_{0,1} \rho$. 
We note that the results in the paper also apply to the case of weak measurements, as discussed in Appendix~\ref{app:weak_measurement}.

Measurements in realistic quantum devices also have readout errors modeled by quantum channels acting on the ancillary measurement device. 
In our dynamics, such errors occur with probability $q$ and can be described by a bit-flip channel before measuring the ancilla:
\begin{align}
    \calN_M[\rho_M] = (1 - q)\rho_M + q X_M \rho_M X_M.
\end{align}

Here, instead of considering individual states after measurements, we keep track of the ensemble (i.e.\ an incoherent mixture) of the post-measurement states on the system and ancilla, which can be obtained by applying complete dephasing channels on the ancillae in the measurement basis.
This ensemble contains, on average, the recoverable information.
However, any code state subject to Pauli errors and strong syndrome measurements always remains block-diagonal in the measurement basis; the complete dephasing does not change the density matrix at all and therefore can be neglected.
The ensemble generated by noisy measurements is then given by the quantum channel
\begin{align}
    \calM[\rho] = \calN_M[U_{Q\to QM} \,\rho\, U_{Q\to QM}^\dagger],\label{eq:meas_channel}
\end{align}
which we refer to as the noisy measurement channel.

Having specified the quantum memory dynamics, we can write down the evolution of the density matrix
\begin{align}
    \rho_{Q\mathbf{M}_t} &= \calM_t \circ \calN[\rho_{Q\mathbf{M}_{t-1}} ] \nonumber\\
    &= \sum_{\bm{m}_t} \calP_{\bm{m}_t} \calN[\rho_{Q\mathbf{M}_{t-1}} ] \calP_{\bm{m}_t} \otimes \calN_{M_t} \left[ \ketbra{\bm{m}_t}{\bm{m}_t}\right],\label{eq:rhoQM_evol}
\end{align}
where $\calM_t$ is the measurement channel for measuring all the stabilizers at time $t$, $\calN_{M_t}$ is the measurement error channel acting on the ancillae introduced at time $t$, and $\calP_{\bm{m}_t}$ is the projector associated with the collection of measurement outcomes $\bm{m}_t$.
Here, we use $\mathbf{M}_t = \{M_1, M_2, \dots, M_t\}$ to denote all the ancilla qubits introduced up to step $t$.

Before proceeding, we introduce a notation for the Pauli operator $\calO_{\bm{a}}$ in terms of a $2N$-component binary vector $\bm{a} = (\bm{a}^x, \bm{a}^z)$,
\begin{align}
    \calO_{\bm{a}} = \ri^{\bm{a}^x\cdot \bm{a}^z} X^{\bm{a}^x} Z^{\bm{a}^z} := \ri^{\bm{a}^x\cdot \bm{a}^z} \prod_{r = 1}^N X_r^{a^x_{r}} Z_r^{a^z_r}.\label{eq:calO_defini}
\end{align}

In this way, we can label the stabilizers using the binary vectors.
We introduce binary vectors $\bm{a}_i$ to label the check operators $g_i = \calO_{\bm{a}_i}$.
The set of vectors $\{\bm{a}_i\}$ for check operators unambiguously specifies the stabilizer code.
Then, an arbitrary stabilizer $g^{\bm{u}}\in\calS$ may be denoted using an $I$-component binary vector $\bm{u}$,
\begin{align}
    g^{\bm{u}} := \prod_{i=1}^I g_i^{u_i}.\label{eq:binary_stabilizer}
\end{align}
If the set of check operators $\calC$ is overcomplete, the redundancies are labeled by $g^{\bm{u}} = \mathds{1}$ for $\bm{u}\neq 0$, and the redundancy group is $\calR = \{\bm{u} : g^{\bm{u}} = \mathds{1}\}$.

We also introduce binary vectors $\bar{\bm{a}}_k$ for $k = 1, \dots, K$ ($k = K+1, \dots, 2K$) to label the logical Pauli-X(Z) operator $\bar{g}_k^{x} = \calO_{\bar{\bm{a}}_k}$ ($\bar{g}_k^{z} = \calO_{\bar{\bm{a}}_{k+K}}$).
We can denote any Pauli operator in the logical space using a $2K$-component binary vector $\bm{\kappa}$,
\begin{align}
    \bar{g}^{\bm{\kappa}} := \prod_{k = 1}^K \ri^{\kappa_k\kappa_{k+K}}(\bar{g}^x_k)^{\kappa_k} (\bar{g}^z_k)^{\kappa_{k+K}}.\label{eq:binary_logical_paulis}
\end{align}

\section{Information diagnostics}\label{sec:diagnostics}
In this section, we introduce coherent information (Sec.~\ref{sec:setup_coherent_information}) and relative entropy (Sec.~\ref{sec:setup_relative_entropy}) as diagnostics of information dynamics in quantum memories.
In Sec.~\ref{sec:setup_fidelity}, we discuss the relation between coherent information and the fidelity of the optimal decoder.

\subsection{Coherent information}\label{sec:setup_coherent_information}
A central quantity to characterize information dynamics is the \emph{coherent information}, which measures the amount of recoverable information in the quantum memory together with the measurement outcomes~\cite{Schumacher:1996dy,Schumacher:2001,preskill1999lecture,wilde2013quantum}.

To formulate this quantity, one can consider the stabilizer code initialized in a maximally entangled Bell state with a reference $R$.
The amount of recoverable information in the system $Q$ together with the measurement record $\bfM := \bfM_T$ is given by the recoverable entanglement with the reference $R$,
\begin{align}
    I_c(R \rangle Q\mathbf{M}) := S(\rho_{Q\mathbf{M}}) - S(\rho_{Q\mathbf{M}R}),
\end{align}
where $S(\rho_{Q\mathbf{M}(R)}) := -\tr\rho_{Q\bfM(R)}\log \rho_{Q\bfM(R)}$ is the von Neumann entropy.
We note that the coherent information retained in an ensemble of quantum trajectories (i.e.\ quantum states labeled by the history of measurement outcomes) is also studied in the context of measurement-induced transitions~\cite{choi2020quantum}.

Analytically computing coherent information is challenging. Here, we consider the R\'enyi version of the coherent information
\begin{align}
    I_c^{(n)}(R\rangle Q\mathbf{M}) = \frac{1}{1-n}\log\frac{\tr \rho_{Q\mathbf{M}}^n}{\tr \rho_{Q\mathbf{M}R}^n},\label{eq:renyi-Ic}
\end{align}
which recovers the coherent information in the limit ${n \to 1}$.

We remark that the R\'enyi version of the coherent information in Eq.~\eqref{eq:renyi-Ic} is different from the average R\'enyi coherent information over measurement trajectories.
Specifically, one can express $I_c^{(n)}$ in terms of the $n$-th moment in each trajectory $\bm{m}$ as
\begin{align}
    I_c^{(n)}(R\rangle Q\mathbf{M}) = \frac{1}{1-n} \log \frac{\sum_{\bm{m}} p_{\bm{m}}^n \tr\rho_{Q,{\bm{m}}}^n}{\sum_{\bm{m}} p_{\bm{m}}^n \tr\rho_{QR,{\bm{m}}}^n}.
\end{align}
Since the averages over ${\bm{m}}$ occur inside the logarithm and are weighted by the $n$-th power of the trajectory probability $p_{\bm{m}}$, this quantity is distinct from the average of the R\'enyi coherent information over trajectories.
Across a decoding transition in quantum memory (e.g. as in Appendix~\ref{app:surface_code}), we expect the two quantities to exhibit different critical points and universalities for $n > 1$.
The two quantities coincide only when $n \to 1$.

\subsection{Relative entropy}\label{sec:setup_relative_entropy}
Intuitively, retaining encoded information requires the final state of the stabilizer code to depend on its initial state.
We thus consider initializing the stabilizer code in distinct initial states and use the distinguishability between their corresponding final states, measured by the \emph{relative entropy}, to characterize the information flow.

First, we consider the quantum relative entropy between the density matrices of the final states of the system $Q$ together with the measurement record $\bfM$.
Here, we prepare the stabilizer code in two orthogonal states, a logical state $\ket{\Psi}$ and an excited state $\ket{\Psi_{\bm{s}}} = w_{\bm{s}}|\Psi\rangle$, where $w_{\bm{s}}$ is an error operator that creates the syndrome $\bm{s} \in \bbZ_2^I$\footnote{The logical (the excited) state need not be a pure state and can be also a mixed state describing the incoherent superposition of different logical states (excited states with the same syndrome $\bm{s}$).}. 
The relative entropy of the two corresponding final states $\rho_{Q\bfM}$ and $\rho_{Q\bfM,\bm{s}}$ is given by
\begin{align}
    D_{\bm{s}} := \tr\rho_{Q\bfM}\log\rho_{Q\bfM} - \tr \rho_{Q\bfM}\log\rho_{Q\bfM,\bm{s}}.
\end{align}
To compute the relative entropy, one can again formulate it as the $n \to 1$ limit of the R\'enyi relative entropies
\begin{align}
    D^{(n)}_{\bm{s}} = \frac{1}{n-1}\log\frac{\tr\rho_{Q\bfM}^n}{\tr\rho_{Q\bfM}\rho^{n-1}_{Q\bfM,\bm{s}}}.
\end{align}

Alternatively, one can consider the relative entropy between two density matrices of the measurement record $\bfM$ only. 
This concerns whether the measurement record can correctly identify the syndromes, which is required for successful quantum error correction.
The density matrix $\rho_\bfM$ is diagonal and represents the probability distribution of measurement outcomes, i.e.\ $\rho_\bfM = \sum_{\bm{m}} p_{\bm{m}} \ketbra{{\bm{m}}}$. 
The relative entropy then reduces to the Kullback–Leibler (KL) divergence between two distributions
\begin{align}
    D_{\mathrm{KL},\bm{s}} := \sum_{\bm{m}} p_{\bm{m}} \log p_{\bm{m}} - p_{\bm{m}} \log p_{{\bm{m}},\bm{s}},
\end{align}
where $p_{{\bm{m}},\bm{s}}$ denotes the probability of trajectory ${\bm{m}}$ when initializing the code in the excited state $\ket{\Psi_{\bm{s}}}$.
The KL divergence can be similarly expressed as the $n \to 1$ limit of the R\'enyi sequence
\begin{align}
    D_{\mathrm{KL},\bm{s}}^{(n)} := \frac{1}{n-1} \log \frac{\sum_{\bm{m}} p_{\bm{m}}^n}{\sum_{\bm{m}} p_{\bm{m}} p_{{\bm{m}},\bm{s}}^{n-1}}.
\end{align}

We note that both the quantum relative entropy $D_{\bm{s}}$ and the KL divergence $D_{\mathrm{KL},\bm{s}}$ are monotones under completely positive trace-preserving maps~\cite{lieb1973proof,araki1975relative,lindblad1975completely,uhlmann1977relative}, and therefore can only decrease in the dynamics considered in this work.
As we show in Sec.~\ref{sec:stat-mech_stabilizer}, the two quantities map to correlation functions in the stat-mech model with different boundary conditions and can detect the potential transition corresponding to the information loss as demonstrated in Appendix~\ref{app:surface_code}. 
Both quantities have been studied in many-body settings; the quantum relative entropy is used to probe the decoding transitions in the decohered toric code~\cite{fan2023diagnostics,lee2023quantum}, and the KL divergence between measurement records is used to diagnose measurement-induced transitions~\cite{bao2020theory}.

\subsection{Fidelity of the optimal decoder}\label{sec:setup_fidelity}
Previous works have characterized the information dynamics in quantum memories using the performance of decoding algorithms.
In particular, extensive work has been done to compute the fidelity of the \emph{maximum likelihood} (ML) decoder for various decohered quantum codes and to use this as a benchmark of the information encoding, both with and without errors in syndrome measurements~\cite{dennis2002topological, bravyi2014efficient, Chubb:2021htd, Bombin:2012jk, iyer2015hardness}.
In what follows, we first briefly review the ML decoder and then demonstrate that the ML decoder is asymptotically optimal for Pauli errors by establishing a relation between its fidelity and coherent information.

To evaluate the decoding fidelity, it is convenient to expand the density matrix $\rho_{Q\bfM}$ as an incoherent superposition of states with different error configurations
\begin{align}
    \rho_{Q\bfM} = \sum_{\{\bm{b}_t, \bm{\epsilon}_t\}} \Pr(\{\bm{b}_t, \bm{\epsilon}_t\}) &\calO_{\Sigma\bm{b}_T}\rho_0 \calO_{\Sigma\bm{b}_T} \nonumber \\
    &\otimes_t \ketbra{\bm{m}_t+\bm{\epsilon}_t}.\label{eq:error_conf_expansion}
\end{align}
where $\bm{b}_t = (\bm{b}^x_t, \bm{b}^z_t)$ denotes the Pauli error applied on $Q$, the binary vector $\bm{\epsilon}_t$ denotes syndrome measurements that are erroneously flipped at time step $t$, and $\Sigma\bm{b}_t := \sum_{\tau=1}^t \bm{b}_\tau$ is the cumulative Pauli error.
Here, we can omit the projector $\calP_{\bm{m}_t}$ at each step in Eq.~\eqref{eq:rhoQM_evol} because the syndrome $\bm{m}_t$ is completely determined by the accumulated Pauli error $\calO_{\Sigma\bm{b}_T}$.
The probability $\Pr(\{\bm{b}_t, \bm{\epsilon}_t\})$ for a given error configuration is a positive number determined by the error rates $p_\alpha$ and $q$.

In a decoding algorithm, our goal is to remove the excitations created by the cumulative error operator $\calO_{\Sigma \bm{b}_T}$ based on the noisy measurement record $\{\bm{m}'_t := \bm{m}_t + \bm{\epsilon}_t\}$.
Here, we assume that the last round of measurement is perfect, i.e.\  $\bm{\epsilon}_T = 0$, and the record contains the genuine measurement outcome $\bm{m}_T$ of the check operators in the final state.
In this case, one can restore a logical state by applying a recovery Pauli operator to completely remove the excitations.
However, the recovery operator is not unique, and applying a recovery operator that differs from the cumulative error by a logical operator $\bar{g}^{\bm \kappa}$ may result in a logical error.
In the maximum likelihood decoder, one evaluates the probability $\Pr(\bm{\kappa} | \bm{m}_T, \{\bm{m'}_t\})$  of cumulative error operators in the class labeled by $\bm{\kappa}$ and selects the recovery operator in the class $\bm{\kappa}^{\textrm{ML}}$ with the maximum probability among $2^{2K}$ classes
\begin{equation}
    \bm{\kappa}^{\textrm{ML}}(\bm{m}_T,\{\bm{m'}_t\}) := \textrm{argmax}_{\bm{\kappa}} \Pr(\bm{\kappa} | \bm{m}_T, \{\bm{m'}_t\}).
\end{equation}
It follows that the ML decoder's decoding failure probability for a given syndrome is
\begin{equation}
    \Delta(\bm{m}_T,\{\bm{m'}_t\}) := 1 - \textrm{max}_{\bm{\kappa}} \Pr(\bm{\kappa} | \bm{m}_T, \{\bm{m'}_t\}).
\end{equation}
In Appendix~\ref{app:ML_infidelity}, we show that the average failure probability $\bar{\Delta} := \overline{\Delta(\bm{m}_T,\{\bm{m'}_t\})}$ has an upper bound:
\begin{equation}
    \bar{\Delta} \leq H(\bm{\kappa}|\bm{m}_T, \{\bm{m'}_t\}),
\end{equation}
where $H(\bm{\kappa}|\bm{m}_T, \{\bm{m'}_t\})$ is the conditional Shannon entropy.
A similar bound was derived in Ref.~\cite{zhao2023extracting} for the case without measurement errors.

In the next step, we relate the Shannon entropy to the coherent information.
To this end, we rewrite the density matrices $\rho_{Q\bfM}$ and $\rho_{Q\bfM R}$ in the error configuration expansion in terms of the syndromes $\bm{m}_T$ and $\bm{m}'_t$
\begin{align}\label{eq:syndrome_expansion}
&\rho_{Q\bfM} = \sum_{\bm{m}_T,\{\bm{m}'_t\}} \frac{1}{2^K} \Pr(\bm{m}_T,\{\bm{m}'_t\}) \calP_{\bm{m}_T}\otimes_t \ketbra{\bm{m}'_t},\nonumber\\
&\begin{aligned}
\rho_{Q\bfM R} = \sum_{\bm{\kappa},\bm{m}_T,\{\bm{m}'_t\}} &\Pr(\bm{\kappa},\bm{m}_T,\{\bm{m}'_t\}) \\
\calO_{0}\bar{g}^{\bm{\kappa}}&\ketbra{\Phi}_{RQ}\bar{g}^{\bm{\kappa}}\calO_{0} \otimes_t \ketbra{\bm{m}'_t},
\end{aligned}
\end{align}
where $\Pr(\bm{m}_T,\{\bm{m}'_t\})$ ($\Pr(\bm{\kappa},\bm{m}_T,\{\bm{m}'_t\})$) is the total probability of error operators that are consistent with the syndrome (and are in the class $\bar{g}^{\bm{\kappa}}$).
Here, we decompose the cumulative error as $\calO_{\Sigma \bm{b}_T} := \calO_0 \bar{g}^\sfc$, where $\calO_0$ is a reference error operator that creates the syndrome $\bm{m}_T$, and $\ket{\Phi}_{RQ}$ is the maximally entangled state between the reference $R$ and the system $Q$, which contains $K$ Bell pairs.
It follows that
\begin{align}
S(\rho_{Q\bfM}) &= H(\bm{m}_T, \{\bm{m}'_t\}) + K\log 2,\\
S(\rho_{Q\bfM R}) &= H(\bm{\kappa},\bm{m}_T,\{\bm{m}'_t\}).
\end{align}
Thus, we have the relation
\begin{align}
    \bar{\Delta} \leq H(\bm{\kappa}|\bm{m}_T,\{\bm{m}'_t\}) = K\log 2 - I_c(R\rangle Q\bfM). \label{eq:Ic_ml_decoder}
\end{align}

In conclusion, we have shown that the ML decoder is asymptotically optimal, i.e.\ it has a vanishing failure probability as long as the coherent information approaches its maximum in the thermodynamic limit.
We remark that the proof works for decoherence described by Pauli channels, and it does not apply to cases with coherent errors. 

In Sec.~\ref{sec:stat-mech}, we show that the stat-mech model derived for coherent information is dual to that obtained for the optimal decoding algorithm, which is consistent with the rigorous relation shown in this section.

\section{Statistical mechanics model}\label{sec:stat-mech}

In this section, we develop stat-mech descriptions for the R\'enyi versions of information diagnostics in our quantum memory dynamics.
The key step is to map $\tr\rho_{Q\bfM}^n$ to the partition function of a stat-mech model.
We then find that the $n$-th moments of different density matrices map to the same model but with different boundary conditions or defect insertions. 
Furthermore, the R\'enyi versions of the information diagnostics are identified with distinct observables in the stat-mech model.

In Sec.~\ref{sec:stat-mech_stabilizer}, we introduce the stabilizer expansion of $\rho_{Q\bfM}$ and use it to obtain a stat-mech model consisting of $n-1$ flavors of Ising spins.
The model exhibits symmetries determined by the redundancies $\calR$ in the check operators.
For example, in the surface code the check operators exhibit a global $\mathbb{Z}_2$ redundancy and the decoding transition is associated with a $\mathbb{Z}_2$ symmetry-breaking transition in the stat-mech model.
In this case, the measurements of check operators suppress the ferromagnetic ordering and therefore increase the decoherence threshold.

Our stat-mech model is dual to the model obtained for the maximum likelihood decoder~\cite{dennis2002topological,Chubb:2021htd,li2024perturbative}.
In Sec.~\ref{sec:stat-mech_error_configuration}, we consider an alternative expansion of the same density matrix in terms of error configurations.
We then obtain a stat-mech model with the same partition function as (i.e.\ dual to) the model in the stabilizer expansion.
We show that our stat-mech model in the error configuration expansion reduces to the one associated with the decoding fidelity of the ML decoder in the limit $n \to 1$.

We remark that the mapping to stat-mech models based on the two expansions of the density matrix and the duality between the resulting models are first proposed in Ref.~\cite{fan2023diagnostics} to study the information diagnostics in topological quantum memory subject to local decoherence.
Our stabilizer expansion generalizes the ``loop expansion" in Ref.~\cite{fan2023diagnostics} and allows better identification of the symmetry in the resulting stat-mech model.

\subsection{Stat-mech models in the stabilizer expansion}\label{sec:stat-mech_stabilizer}
Developing stat-mech models in the stabilizer expansion takes two steps. 
In Sec.~\ref{sec:stat-mech_stabilizer_dynamics}, we express the density matrix at time $T$ as a sum of operators weighted by real positive numbers.
Then, in Sec.~\ref{sec:stat-mech_stabilizer_quantities}, we map the information diagnostics, the coherent information and relative entropy, to the partition functions of stat-mech models.

\subsubsection{State evolution under noise and imperfect measurements}\label{sec:stat-mech_stabilizer_dynamics}
Here, we formulate the evolution of the density matrix $\rho_{Q\bfM}$ under $T$ layers of decoherence and noisy syndrome measurements in the stabilizer expansion.
We consider the initial state $\rho_{Q\bfM_0}$ to be the maximally mixed state in the code space of a stabilizer code $\calS$
\begin{align}
    \rho_{Q\bfM_0} = \frac{1}{2^K} \prod_{i \in I}\frac{1 + g_i}{2} = \frac{1}{2^N} \sum_{\bm{u}} g^{\bm{u}},
\end{align}
where $\bm{u} \in \bbZ_2^I$ is an $I$-component binary vector that labels the stabilizer $g^{\bm{u}} \in \calS$ (as in Eq.~\eqref{eq:binary_stabilizer}).
From now on, we do not keep track of the overall prefactor in the expansion, which can always be restored by enforcing $\tr\rho = 1$.

The evolution of $\rho_{Q\bfM_0}$ under physical decoherence takes a simple form in the stabilizer expansion because Pauli operators acquire only overall prefactors under Pauli channels, i.e.\
\begin{equation}
\calN[\rho_{Q\bfM_0}] = \sum_{\bm{u}} e^{-H(\bm{u})}g^{\bm{u}}.
\end{equation}
To explicitly write down the prefactor, we express the stabilizer as
\begin{align}
    g_i = \ri^{\bm{a}_i^x\cdot \bm{a}_i^z} X^{\bm{a}_i^x} Z^{\bm{a}_i^z},
\end{align}
where $\bm{a}_i^{x(z)}$ is a binary vector of length $N$ specifying the support of Pauli-X(Z) on physical qubits.

The prefactor then takes the form
\begin{multline}\label{eq:changed_weight}
e^{-H(\bm{u})} \\ 
= \prod_{r} \bigg[\bigg(1-\sum_{\alpha}p_\alpha\bigg) + p_xh_r^z+ p_zh_r^x + p_yh_r^xh_r^z\bigg],
\end{multline}
where the binary variable $h_r^{x(z)} := \prod_i (-1)^{u_i a_{i,r}^{x(z)}}$ tracks the support of Pauli-X(Z) at site $r$.
This prefactor can be viewed as the Boltzmann weight of a classical Hamiltonian
\begin{equation}\label{eq:changed_H}
H(\bm{u}) = \sum_r -J_z h_r^x - J_x h_r^z - J_y h_r^xh_r^z
\end{equation}
up to a constant term independent of $\bm{u}$, with
\begin{equation}\label{eq:mu_defs}
    J_{\alpha_1} = -\frac{1}{4}\log\frac{(1-2p_{\alpha_1}-2p_{\alpha_2})(1-2p_{\alpha_1}-2p_{\alpha_3})}{1-2p_{\alpha_2}-2p_{\alpha_3}},
\end{equation}
where $\alpha_{1,2,3}$ are different labels chosen from $\{x,y,z\}$.

Next, we obtain the evolution of $\rho_{Q\bfM}$ under noisy syndrome measurements.
Each Pauli operator $g^{\bm{u}}$ in the stabilizer expansion is evolved under the measurement channel $\calM_{t,i}$ associated with measuring $g_i$ at time $t$, which takes the form
\begin{align}
    \calM_{t,i}[g^{\bm{u}}] &= \sum_{m_{t,i}=0,1} \calP_{m_{t,i}} g^{\bm{u}} \calP_{m_{t,i}} \calN_{M}[\ketbra{m_{t,i}}] \nonumber \\
    &= \frac{1}{2} g^{\bm{u}} \left(1 + e^{-\mu_q} g_i Z_{t,i}\right),
\end{align}
where $\calP_{m_{t,i}} = (1 + (-1)^{m_{t,i}} g_i)/2$ is the projector onto corresponding eigenstates, $\mu_q = -\log(1-2q)$, and $Z_{t,i}$ is the Pauli-Z operator for the measurement ancilla.
After a full layer of stabilizer measurements at the first time step, the resulting density matrix is
\begin{align}
    \rho_{Q\bfM_1} = \sum_{\bm{u}_0,\bm{u}_1}e^{-H(\bm{u}_0)-\mu_q|\bm{u}_1-\bm{u}_0|}g^{\bm{u}_1} Z_1^{\bm{u}_1-\bm{u}_0},
\end{align}
where $|\bm{u}| = \sum_i u_{i}$ counts the number of nonzero elements in the binary vector $\bm{u}$, and $Z_1^{\bm{u}} := \otimes_{i=1}^I Z_{1,i}^{u_i}$ is a product of Pauli-Z on ancillae. We remark that $|\bm{u}_1-\bm{u}_0|$ serves as an Ising coupling between spins at consecutive times.

After $T$ layers of decoherence and noisy measurements, the density matrix $\rho_{Q\bfM}$ takes the form
\begin{align}\label{eq:stabilizer_expansion}
    \rho_{Q\bfM} = \sum_{\{\bm{u}_t\}} e^{-\calH(\{\bm{u}_t\})} g^{\bm{u}_T}\otimes_{t =1}^T Z_t^{\bm{u}_t-\bm{u}_{t-1}},
\end{align}
where the weight of the Pauli operator can be written as the Boltzmann weight associated with a classical Hamiltonian
\begin{align}
\calH(\{\bm{u}_t\}) = \sum_{t = 1}^{T} H(\bm{u}_{t-1}) + \mu_q|\bm{u}_t-\bm{u}_{t-1}|.
\end{align}
Here, the Hamiltonian involves $T+1$ layers of classical variables with intralayer coupling described by $H(\bm{u}_{t})$ in the first $T$ layers and interlayer coupling $\mu_q|\bm{u}_t-\bm{u}_{t-1}|$ controlled by the measurement error rate.
We note that in the case that $\calS$ has redundancies, a different set of binary vectors $\{\bm{u}'_t\}$ related to $\{\bm{u}_t\}$ via a linear transformation can represent the same Pauli operator $g^{\bm{u}_t}$. 
Here, all $\bm{u}_t$ must transform together to keep the Pauli-Z operator on the ancillae, i.e.\ $|\bm{u}_t -\bm{u}_{t-1}|$, invariant. 
The transformation, therefore, relates two sets of binary vectors with the same weight and is a symmetry of the Hamiltonian.
This indicates that the Hamiltonian $\calH(\{\bm{u}_t\})$ exhibits a symmetry group $\calR$ given by the redundancy of the stabilizers.
In addition, we note that the Hamiltonian can exhibit ``spacetime" symmetries arising from the symmetries of the code and decoherence model under translations, rotations, etc.

\emph{Examples.---}
We here discuss the Hamiltonian $\calH(\bm{u})$ in a few examples.
In the 2D toric code (the surface code on the 2D square lattice with periodic boundary conditions), the qubits are defined on the edges, and the model involves classical variables associated with check operators on the vertices and plaquettes.
The Pauli-X(Z) operator for each qubit is only involved in two check operators.
Accordingly, $h_r^{x(z)}$ in the Hamiltonian $H(\bm{u})$ is a two-body term, and $h_r^x h_r^z$ is a four-body term.
The check operators in the toric code have redundancies as the product of all plaquette (star) check operators is the identity, indicating an $\calR = \bbZ_2 \times \bbZ_2$ symmetry of the Hamiltonian.

In contrast, the 2D planar surface code with open boundary conditions does not have extraneous check operators. 
This indicates that the Hamiltonian does not possess an exact $\calR$ symmetry.
Nevertheless, the product of all plaquette (star) check operators in the planar code cancels in the bulk with non-trivial support only on the boundary, indicating an approximate symmetry that is broken only on the boundary.
We note that such approximate symmetry is still crucial for governing the potential symmetry-breaking transition in the thermodynamic limit.

In the 2D repetition code on a square lattice with periodic boundary conditions, each qubit (defined on vertices) is involved in four $ZZ$-check operators, leading to a four-body coupling $h_r^x$ in the Hamiltonian.
Due to the local redundancies in the check operators, namely that the product of four checks around a plaquette is the identity, the Hamiltonian exhibits an extensive number of $\bbZ_2$ symmetries, whose local structure allows one to identify $H(\bm{u})$ as a $\bbZ_2$ gauge theory.

\subsubsection{Stat-mech models for information diagnostics}\label{sec:stat-mech_stabilizer_quantities}
We are now ready to develop stat-mech models for coherent information, decoding fidelity, and relative entropy. 
The key step is to identify the $n$-th moment of the density matrix, e.g. $\tr\rho_{Q\bfM}^n$, with the partition function of a $(n-1)$-flavor Ising model,
\begin{equation}
    \calZ_n := \tr \rho_{Q\bfM}^n.
\end{equation}

In the stabilizer expansion, the $n$-th moment $\tr \rho_{Q\bfM}^n = \calZ_n$ involves a factor
\begin{align}
    \Omega(\{\bm{u}^{\sfa}\}) :=& \tr(\prod_{\sfa=1}^n g^{\bm{u}_{T}^{\sfa}}\prod_{t=1}^T Z_t^{\bm{u}_t^{\sfa}-\bm{u}_{t-1}^{\sfa}} ) \nonumber \\
    =& \sum_{\bm{r} \in \calR}\prod_{t = 0}^{T}\delta\left(\sum_{\sfa=1}^n \bm{u}_t^\sfa = \bm{r}\right),\label{eq:delta_functions}
\end{align}
where $\bm{u}^\sfa$ is the binary variable associated with the $\sfa$-th copy of the density matrix $\rho_{Q\bfM}$.
Here, $\Omega$ is nonvanishing only if $\sum_\sfa \bm{u}_t^\sfa \in \calR$ for all $0\leq t \leq T$.
We note that we have omitted an overall constant in the second equality.

The delta function in Eq.~\eqref{eq:delta_functions} allows one to write the $n$-th moment as a partition function of classical variables in the first $n-1$ copies
\begin{align}
\calZ_n = \sum_{\bm{r} \in \calR} \sum_{\{\bm{u}_t^\sfa\}} e^{-\sum_{\sfa = 1}^{n-1} \calH(\{\bm{u}_t^\sfa\}) - \calH(\{\sum_{\sfa = 1}^{n-1}\bm{u}_t^\sfa + \bm{r}\})}.
\end{align}
We note that since $\calH(\{\bm{u}_t + \bm{r}\}) = \calH(\{\bm{u}_t\})$ for $\bm{r} \in \calR$, the summation over the redundancy group gives only an overall prefactor and therefore can be omitted.

It is convenient to introduce classical Ising spin variables $\sigma^\sfa_{t,i} := (-1)^{u_{t,i}^\sfa}$ to re-express the partition function as
\begin{equation}\label{eq:stabilizer_expansion_Zn}
    \calZ_n = \sum_{\bm{\sigma}^{1},\dots,\bm{\sigma}^{n-1}} e^{-\calH^{(n)}(\{\bm{\sigma}^{\sfa}\})}
\end{equation}
with $\calH^{(n)}$ being the Hamiltonian of $(n-1)$-flavor Ising spins,
\begin{equation}\label{eq:stabilizer_expansion_flavor_hamiltonian}
    \calH^{(n)}(\{\bm{\sigma}^{\sfa}\}) := \sum_{\sfa=1}^{n-1}\calH(\bm{\sigma}^{\sfa}) + \calH\left(\prod_{\sfa=1}^{n-1}\bm{\sigma}^{\sfa}\right).
\end{equation}
We note that the Hamiltonian $\calH^{(n)}$ has open boundary conditions on the final $t=T$ layer. 

The symmetry of $\calH^{(n)}$ is convenient to see by formulating the Hamiltonian  $\calH^{(n)}$ as $n$ copies of single-flavor Hamiltonian $\calH(\bm{\sigma}^\sfa)$ subject to the constraint $\bm{\sigma}^n = \prod_{\sfa = 1}^{n-1}\bm{\sigma}^\sfa$.
Then, the symmetry of $\calH^{(n)}$ contains $n$ copies of the symmetry $\calR$ of the single-flavor Hamiltonian.
The $\calR^{n}$ symmetry is further extended by the $S_n$ symmetry permuting over $n$ identical copies of $\calH(\bm{\sigma}^\sfa)$.
After factoring out a copy of symmetry $\calR$ due to the constraint, we obtain the global symmetry $(\calR^{n}\rtimes S_n)/\calR$ of the Hamiltonian $\calH^{(n)}$.
Again, we note that when evaluating the thermodynamic properties in the stat-mech model, one also needs to involve the approximate symmetries and the spacetime symmetries as a part of $\calR$.

We remark that the stat-mech model reduces to a single-flavor model for $n = 2$ and $n \to \infty$. 
At $n = 2$, the Hamiltonian $\calH^{(2)} = 2\calH(\bm{\sigma}^1)$ contains a single flavor.
At $n \to \infty$, the interaction $\calH(\prod_\sfa \bm{\sigma}^\sfa)$ between flavors becomes negligible\footnote{One can view the $n-1$ flavors of the Ising spin as an additional dimension. Let $\bm{\tau}^\sfa := \prod_{\sfb = \sfa}^{n-1}\bm{\sigma}^\sfb$. The Hamiltonian is $\calH^{(n)} = \calH(\bm{\tau}^1) + \sum_{\sfa=1}^{n-2}\calH(\bm{\tau}^{\sfa}\bm{\tau}^{\sfa+1})$. The inter-replica coupling becomes $\calH(\bm{\tau}^1)$, a boundary field in the flavor dimension, which can be neglected in the limit $n \to \infty$.}, and the system consists of $n-1$ independent copies of $\calH(\bm{\sigma}^\sfa)$.
These two cases provide useful intuition regarding the potential phases and phase transitions in the $n \to 1$ limit, despite different universalities due to different symmetries of the model.

\emph{Coherent information.---}
To calculate the R\'enyi coherent information, we now map the $n$-th moment $\tr\rho_{Q\bfM R}^n$ to the stat-mech model and show that the coherent information corresponds to the excess free energy of inserting defects in the stat-mech model.

To begin, the initial state $\rho_{Q\bfM_0 R}$ is the maximally entangled state between the code space of the system $Q$ and the reference $R$ that contains $K$ qubits
\begin{align}
\rho_{Q\bfM_0 R} &= \prod_{k=1}^K\frac{1 + s^x_k \bar{g}^x_k}{2}\frac{1 + s^z_k \bar{g}^z_k}{2}\prod_{i=1}^I \frac{1+g_i}{2} \nonumber \\
&= \frac{1}{2^{K+N}}\sum_{\bm{u},\bm{\kappa}}g^{\bm{u}} \bar{g}^{\bm{\kappa}} s^{\bm{\kappa}} ,
\end{align}
where $s^{x(z)}_k$ is the Pauli-X(Z) operator for the $k$-th qubit in the reference $R$. The maximally entangled state is stabilized by $s^x_k\bar{g}^x_k$ and $s^z_k\bar{g}^z_k$.
Each term in the stabilizer expansion can be labeled by the binary vector $\bm{u}$ together with a $2K$-component binary vector $\bm{\kappa}$ which labels the content in the logical space and reference. 
We will omit the constant prefactor in the second equality in what follows.

Under the time evolution of quantum memory, the density matrix $\rho_{Q\bfM R}$ evolves as
\begin{align}
    \rho_{Q\bfM R} = \sum_{\{\bm{u}_t\},\bm{\kappa}} e^{-\calH_{\bm{\kappa}}(\{\bm{u}_t\})} g^{\bm{u}_T} \bar{g}^{\bm{\kappa}} s^{\bm{\kappa}} \otimes_{t = 1}^T Z_t^{\bm{u}_t-\bm{u}_{t-1}}.\label{eq:stabilizer_expansion_coherent_info}
\end{align}
Here, the logical operator $\bar{g}^{\bm{\kappa}}$ and the Pauli operator $s^{\bm{\kappa}}$ on the reference do not evolve, and the terms in the expansion acquire a weight $e^{-\calH_{\bm{\kappa}}}$ that depends on the logical operator $\bm{\kappa}$.

The Hamiltonian that governs the weight is given by
\begin{align}
    \calH_{\bm{\kappa}}(\{\bm{u}_t\}) &= \sum_{t = 1}^T H_{\bm{\kappa}}(\bm{u}_{t-1}) + \mu_q|\bm{u}_t-\bm{u}_{t-1}|, \\
    H_{\bm{\kappa}}(\bm{u}) &= \sum_r-J_zh_{\bm{\kappa},r}^x - J_xh_{\bm{\kappa},r}^z - J_yh_{\bm{\kappa},r}^xh_{\bm{\kappa},r}^z,
\end{align}
where 
\begin{align}
h_{\bm{\kappa},r}^x(\bm{u}) &:= (-1)^{\sum_{k=1}^{2K} \kappa_k \bar{a}_{k,r}^x} h_{r}^x(\bm{u}), \\
h_{\bm{\kappa},r}^z(\bm{u}) &:= (-1)^{\sum_{k=1}^{2K} \kappa_k \bar{a}_{k,r}^z}h_{r}^z(\bm{u}).
\end{align}
Here, $\bar{\bm{a}}_k$ is the $2N$-component binary vector associated with the $2K$ logical (X and Z) operators.
The logical operator $\bar{g}^{\bm{\kappa}}$ in the expansion therefore controls the sign of $h^{x,z}_r$ effectively inserting a defect in the stat-mech model along the temporal direction.

Next, we consider the $n$-th moment $\tr\rho^n_{Q\bfM R}$ in the stabilizer expansion. 
The trace is non-vanishing only if $\bm{\kappa}^n = \sum_{\sfa=1}^{n-1} \bm{\kappa}^\sfa$.
The $n$-th moment thus takes the form
\begin{align}
    \tr\rho^n_{Q\bfM R} = \frac{1}{2^{(n-1)K}}\sum_{\{\bm{\kappa}^\sfa\}} \calZ_n(\{\bm{\kappa}^\sfa\}),
\end{align}
which is a summation of the partition functions with different defect insertions labeled by $\bm{\kappa}^\sfa$:
\begin{align}
    \calZ_n(\{\bm{\kappa}^\sfa\}) = \sum_{\{\bm{u}^\sfa\}} e^{-\sum_{\sfa=1}^{n-1} \calH_{\bm{\kappa}^\sfa}(\bm{u}^\sfa) - \calH_{\sum_\sfa \bm{\kappa}^\sfa}(\sum_\sfa \bm{u}^\sfa)}.
\end{align}

We can thus write down the R\'enyi-$n$ coherent information as
\begin{align}
    I_c^{(n)}(R\rangle Q\bfM) = \frac{1}{n-1}\log \left(\sum_{\{\bm{\kappa}^\sfa\}} e^{-\Delta F(\{\bm{\kappa}^\sfa\})} \right) - K\log 2,\label{eq:renyi-n_coherent_info_stat-mech}
\end{align}
where $\Delta F(\{\bm{\kappa}^\sfa\}) := -\log \calZ_n(\{\bm{\kappa}^\sfa\}) + \log \calZ_n$ is the excess free energy of inserting the defect.

\emph{Relative entropy.---} To evaluate the relative entropy, we first express $\rho_{Q\bfM_0, \bm{s}} = w_{\bm{s}}\rho_{Q\bfM_0} w_{\bm{s}}^\dagger$ in the stabilizer expansion
\begin{align}
\rho_{Q\bfM_0, \bm{s}} = \sum_{\bm{u}} (-1)^{\bm{s}\cdot\bm{u}} g^{\bm{u}},
\end{align}
where we recall that $\bm{s}$ is a nontrivial syndrome, resulting from Pauli operator $w_{\bm{s}}$.

After evolving for $T$ steps, the density matrix $\rho_{Q\bfM,\bm{s}}$ takes the form
\begin{align}
    \rho_{Q\bfM, \bm{s}} = \sum_{\{\bm{u}_t\}} (-1)^{\bm{s}\cdot\bm{u_0}} e^{-\calH(\{\bm{u}_t\})} g^{\bm{u}_T} \otimes_{t = 1}^T Z_t^{\bm{u}_t-\bm{u}_{t-1}}.
\end{align}
Accordingly, the $n$-th moment $\tr\rho_{Q\bfM}\rho^{n-1}_{Q\bfM,s}$ is given by
\begin{align}
\tr\rho_{Q\bfM}\rho^{n-1}_{Q\bfM,\bm{s}} = \sum_{\{\bm{u}^\sfa\}} (-1)^{\bm{s}\cdot\bm{u^1_0}} e^{-\calH^{(n)}(\{\bm{u}^\sfa\})}.
\end{align}
We note that the phase factor is inserted for the last $n-1$ copies, and we have used the constraint $\bm{u}_0^1 = \sum_{\sfa=2}^n \bm{u}_0^\sfa$ to simplify the expression.

We can thus write down the relative entropy as a boundary correlation function between classical spins in the first copy
\begin{align}
    D_{\bm{s}}^{(n)} = -\frac{1}{n-1}\log \left< \prod_i (\sigma^1_{0,i})^{s_i}\right>,\label{eq:stabilizer_expansion_relative_entropy}
\end{align}
where $\langle \cdot \rangle$ denotes the expectation value in the defect-free stat-mech model of $n-1$ flavors, and we recall that $\sigma^{\sfa}_{t,i} = (-1)^{u^\sfa_{t,i}}$.

\emph{KL divergence.---} Last, we consider the R\'enyi-$n$ version of the KL divergence in the stat-mech model.
The key difference is that we trace over the system $Q$ to obtain the reduced density matrix of the measurement ancillae $\rho_{\bfM} = \tr_Q \rho_{Q\bfM}$.
The trace is non-vanishing only if $g^{\bm{u}_T}$ is the identity, i.e $\bm{u}_T \in \calR$.
We note that for any $\bm{u}_T \in \calR$, one can shift $\bm{u}_t \mapsto \bm{u}_t - \bm{u}_T$ while keeping the weight invariant in the stabilizer expansion.
Thus, without loss of generality, we can set $\bm{u}_T = 0$.
Then, the expansion takes the form
\begin{align}
    \rho_{\bfM} = \sum_{\bm{u}_t} e^{-\calH^+(\{\bm{u}_t\})} Z_T^{-\bm{u}_{T-1}} \otimes_{t = 1}^{T-2} Z_t^{\bm{u}_t - \bm{u}_{t-1}},
\end{align}
where
\begin{align}
    \calH^+(\{\bm{u}_t\}) = \sum_{t=1}^T H(\bm{u}_{t-1}) + \sum_{t = 1}^{T-1} \mu_q |\bm{u}_t - \bm{u}_{t-1}| + \mu_q |\bm{u}_{T-1}|.
\end{align}
Here, the last term changes under the symmetry transformation $\bm{u}_t \mapsto \bm{u}_t + \bm{v}$ with $\bm{v}\in \calR$, and therefore is a symmetry-breaking field at the top boundary $t = T$ in the stat-mech model.
Accordingly, the $n$-th moment $\tr\rho_{\bfM}^n$ takes the form
\begin{align}
    \tr\rho_{\bfM}^n &= \sum_{\{\bm{u}^\sfa\}} e^{-\calH^{(n),+}(\{\bm{u}^\sfa\})} \nonumber \\
    &= \sum_{\{\bm{u}^\sfa\}} e^{\sum_{\sfa=1}^{n-1}-\calH^{+}(\bm{u}^\sfa) - \calH^{+}(\sum_{\sfa=1}^{n-1}\bm{u}^\sfa)}.
\end{align}

Similar to the quantum relative entropy, the R\'enyi-$n$ KL divergence is given by
\begin{align}
    D_{\text{KL},\bm{s}}^{(n)} = -\frac{1}{n-1}\log \left< \prod_i (\sigma^1_{0,i})^{s_i} \right>_+,
\end{align}
where $\langle \cdot \rangle_+$ denotes expectation value in the defect-free stat-mech model of $n-1$ flavors with symmetry-breaking boundary conditions.

The existence of symmetry-breaking boundary fields may lead to qualitative distinctions between $D_{\text{KL}}$ and $D$, linked to the subtle difference between successful error correction and retaining quantum information in $\rho_{Q\bfM}$.
In the example of the 2D surface code, the boundary field always orders the spins in the multi-flavor Ising model (derived in Appendix~\ref{app:surface_code}) when $t = O(1)$, leading to $D_{\text{KL}} = O(1)$.
This indicates that noisy measurements cannot identify distinct error syndromes after $O(1)$ time, although the coherent quantum information $I_c(R\rangle Q\bfM)$ may remain maximal for sufficiently small error rates.
This highlights the need for $t = O(L)$ rounds of noisy measurements for successful error correction.
We explain the same physics in the 3D RPIM for the ML decoder in Appendix~\ref{app:boundary_condition}.

\subsection{Stat-mech model in the error configuration expansion}\label{sec:stat-mech_error_configuration}
In this section, we take the error configuration expansion of the mixed density matrix $\rho_{Q\bfM}$ in Eq.~\eqref{eq:error_conf_expansion} and map the $n$-th moment $\tr\rho_{Q\bfM}^n$ to the partition function of a stat-mech model, which is dual to the model obtained in the stabilizer expansion (Eq.~\eqref{eq:stabilizer_expansion_Zn}).
In the limit of $n \to 1$, the stat-mech model in the error configuration expansion reduces to a disordered model on the Nishimori line, which governs the decoding fidelity of the maximum likelihood decoder.

First, we use the error configuration expansion in Eq.~\eqref{eq:error_conf_expansion} to express the $n$-th moment $\tr\rho^n_{Q\bfM}$ as
\begin{align}
    \tr \rho_{Q\bfM}^n = \sum_{\{\bm{b}_t^\sfa, \bm{\epsilon}_t^\sfa\}} &\prod_{\sfa=1}^{n} \Pr(\{\bm{b}_t^\sfa,\bm{\epsilon}_t^\sfa\}) \prod_{t,\sfa} \delta_{\bm{m}_t^\sfa+\bm{\epsilon}_t^{\sfa},\bm{m}_t^{\sfa+1}+\bm{\epsilon}_t^{\sfa+1}} \nonumber \\
    &\tr\left[\prod_{\sfa=1}^n \calO_{\Sigma\bm{b}_T^\sfa}\,\rho_0\,\calO_{\Sigma\bm{b}_T^\sfa}^\dagger\right],\label{eq:nth_error_conf}
\end{align}
where $\Sigma \bm{b}_T^\sfa := \sum_{\tau = 1}^T \bm{b}_\tau^\sfa$ is a shorthand notation, and we again choose $\rho_0$ to be the maximally mixed state in the logical space.
In this expansion, the $n$-th moment becomes a constrained sum over error configurations weighted by their probability.

The error configuration has a non-vanishing weight only if the following two conditions are satisfied: (1) the physical and measurement errors in each copy produce the same error syndromes at every time step $t$, i.e.\
\begin{align}
    \bm{m}_t^1+\bm{\epsilon}_t^1 = \bm{m}_t^\sfa+\bm{\epsilon}_t^{\sfa}, \text{ for } \sfa = 2,3,\dots, n;
\end{align}
(2) the cumulative physical errors $\Sigma \bm{b}_T^\sfa$ produce the same syndromes in each copy, i.e.\
\begin{align}
    \bm{m}_T^1 = \bm{m}_T^{\sfa}, \text{ for } \sfa = 2,3,\dots, n.
\end{align}
Here, $\bm{m}_t^\sfa$ is the true eigenvalue of measured stabilizers and is flipped only by the physical error operators,
\begin{align}
    {m}_{t,i}^\sfa - {m}_{t-1,i}^\sfa = \Langle \bm{a}_i, \bm{b}_t^\sfa\Rangle,\label{eq:m_conf}
\end{align}
where $\Langle \bm{a}, \bm{b}\Rangle = \bm{a}^x\cdot \bm{b}^z + \bm{a}^z\cdot \bm{b}^x$ is the binary symplectic form.
Hence, the first condition requires
\begin{align}
    \Langle \bm{a}_i, \bm{b}^1_{t}\Rangle + {\epsilon}_{t,i}^1-{\epsilon}_{t-1,i}^1 = \Langle \bm{a}_i, \bm{b}^\sfa_{t}\Rangle + {\epsilon}_{t,i}^\sfa-{\epsilon}_{t-1,i}^\sfa.\label{eq:error_relation}
\end{align}
We note that $\bm{m}_0^\sfa = \bm{\epsilon}_0^\sfa = \bm{0}$ in Eqs.~\eqref{eq:m_conf} and~\eqref{eq:error_relation} for $t = 1$.
The second condition requires the same readout errors in the last round of measurements in each copy, i.e.\ $\bm{\epsilon}_T^1 = \bm{\epsilon}_T^\sfa$.

Before proceeding, we provide an intuitive understanding of Eq.~\eqref{eq:error_relation}.
One can view the symplectic form $\Langle \bm{a}_i, \bm{b}_t\Rangle = \partial \bm{b}_t$ as the boundary of the physical error chain $\bm{b}_t$.
The difference in measurement error $\epsilon_{t,i} - \epsilon_{t-1,i}$ is the boundary $\partial \bm{\epsilon}_t$ of measurement error chain $\bm{\epsilon}_t$.
Equation~\eqref{eq:error_relation} then takes the form
\begin{align}
    \partial (\bm{b}_t^1 - \bm{b}_t^\sfa) + \partial (\bm{\epsilon}_t^1 - \bm{\epsilon}_t^\sfa) = 0.
\end{align}
This indicates the (physical and measurement) error configuration in the first and the $\sfa$-th copy form a closed chain.

We can thus introduce the new variables $\bm{e}_t^\sfa$ to specify the closed chains and label the configuration of $(\bm{b}_t^\sfa, \bm{\epsilon}_t^\sfa)$ with non-vanishing weights:
\begin{align}
    \bm{b}_t^\sfa - \bm{b}_t^1 &= \bm{e}_t^\sfa, \label{eq:diff_phy_error}\\
    \delta{\epsilon}_{t,i}^\sfa - \delta{\epsilon}_{t,i}^1 &= \Langle \bm{a}_i, \bm{e}_t^\sfa\Rangle.\label{eq:diff_meas_error}
\end{align}
Because each copy has the same syndrome $\bm{m}_T^1 = \bm{m}_T^\sfa$ at the final step, the cumulative physical errors in different copies can only differ by a stabilizer $\bm{a}_{\bm{v}}^\sfa := \sum_{i=1}^{I_0} v_{i}^\sfa \bm{a}_i$ and a logical operator $\bar{\bm{a}}_{\bm{\kappa}}^\sfa := \sum_{k=1}^{2K}{\kappa}^{\sfa}_{k} \bar{\bm{a}}_k$
\begin{align}
    \Sigma \bm{e}_T^\sfa = \Sigma \bm{b}_T^\sfa - \Sigma \bm{b}_T^1 = \bm{a}^\sfa_{\bm{v}} + \bar{\bm{a}}_{\bm{\kappa}}^\sfa.
\end{align}
Thus, one can divide the error chain $(\bm{b}_t^\sfa, \bm{\epsilon}_t^\sfa)$ in the $\sfa$-th copy into $2^{2K}$ classes according to the equivalence relation of $\Sigma \bm{e}_T^\sfa$ modulo stabilizers $\bm{a}^\sfa_{\bm{v}}$, i.e.\ $[\Sigma \bm{e}_T^\sfa] := \bm{\kappa}$.

With the newly introduced variables, one can express the $n$-th moment as the partition function of a stat-mech model, i.e.\ $\tr\rho_{Q\bfM}^n = \calZ_n$,
\begin{align}
    \calZ_n = \frac{1}{2^{K(n-1)}}\sum_{\bm{b}^1, \bm{\epsilon}^1} \Pr(\bm{b}^1,\bm{\epsilon}^1) \left(\sum_{\bm{\kappa}}\calZ'(\bm{b}^1,\bm{\epsilon}^1,\bm{\kappa})\right)^{n-1},\label{eq:n-1th_disorder_partition_function}
\end{align}
where the prefactor $1/2^{K(n-1)}$ originates from the normalization of $\rho_0$, and
\begin{align}
    &\calZ' (\bm{b}^1,\bm{\epsilon}^1,\bm{\kappa}) \nonumber \\
    &= \sum_{\{\bm{e}_t\}}\Pr\left(\{\bm{b}_t^1+ \bm{e}_t,\epsilon_{t,i}^1+\Langle \bm{a}_i, \Sigma\bm{e}_t\Rangle\}\right)\delta_{[\Sigma \bm{e}_T] =\bm{\kappa}}.\label{eq:disorder_partition_function}
\end{align}
The partition function $\calZ_n$ can be regarded as the disorder average of $n-1$ copies of the disordered partition function.
Specifically, we view $\bm{b}^1$ and $\bm{\epsilon}^1$ as disorder variables.
The disordered stat-mech model involves binary variables $\bm{e}_t$ as the degrees of freedom.
We introduce an additional label $\bm{\kappa}$ for later convenience; it is associated with the decoding class in the QEC problem.
In the example of the toric code, the stat-mech model is a 3D RPIM, and $\bm{\kappa}$ indicates the defect insertions (i.e.\ $\pi$-flux lines) along non-contractible loops in two spatial directions.
We note that the stat-mech model above and the model obtained in the stabilizer expansion in Sec.~\ref{sec:stat-mech_stabilizer} describe the same quantity and therefore are dual.

We remark that the disordered model in Eq.~\eqref{eq:n-1th_disorder_partition_function} has a gauge symmetry.
The Boltzmann weight $\Pr$ in Eq.~\eqref{eq:disorder_partition_function} is invariant under a local change of the disorder configuration
\begin{align}
    \bm{b}_t^1 \mapsto \bm{b}_t^1 + \bm{e}_t^1 , \quad \delta\epsilon_{t,i}^1\mapsto \delta\epsilon_{t,i}^1 + \Langle \bm{a}_i, \bm{e}_t^1\Rangle \label{eq:shift_disorder}
\end{align}
together with a redefinition of the degrees of freedom in the model\footnote{In the 2D RBIM for decoding the toric code with perfect syndrome measurements, the gauge transformation is a spin flip in Eq.~\eqref{eq:spin_flip} combined with a shift in the random bond configuration as in Eq.~\eqref{eq:shift_disorder}.}, 
\begin{align}
    \bm{e}_t \mapsto \bm{e}_t - \bm{e}_t^1.\label{eq:spin_flip}
\end{align} 
Hence, the partition function $\calZ'$ is invariant under the shift in disorder configuration in Eq.~\eqref{eq:shift_disorder}, and its value $\calZ'$ is fully determined by the erroneous syndromes $\bm{m}'_t:= \bm{m}_t+\bm{\epsilon}_t$, true stabilizer eigenvalues $\bm{m}_T$ in the final state, and logical sectors $\bm{\kappa}$.

To understand the stat-mech model in the $n \to 1$ limit, we consider a specific information diagnostic, the coherent information, which according to Eq.~\eqref{eq:Ic_ml_decoder} is related to the Shannon entropy of the distribution over decoding sectors in the limit $n \to 1$. 
In Appendix~\ref{app:decoding_fidelity_error_configuration}, we show that, for each syndrome $(\bm{m}'_t, \bm{m}_T)$, the partition function $\calZ'$ determines the probability of each decoding sector $\bm{\kappa}$ in the ML decoder:
\begin{align}
    P(\bm{\kappa} | \bm{m}'_t, \bm{m}_T) = \frac{\calZ'(\bm{b}^1_0,\bm{\epsilon}^1_0,\bm{\kappa})}{\sum_{\bm{\kappa}}\calZ'(\bm{b}_0^1,\bm{\epsilon}_0^1,\bm{\kappa})},
\end{align}
where $(\bm{b}_0^1, \bm{\epsilon}_0^1)$ is a reference error configuration that produces the syndrome $(\bm{m}'_t, \bm{m}_T)$\footnote{We note that the decoding sector $\bm{\kappa}$ depends on the choice of $(\bm{b}_0^1, \bm{\epsilon}_0^1)$; however, the probability of the most likely sector is independent of this choice.}.
Therefore, the disordered model $\calZ'$ governs the information diagnostics in the limit $n \to 1$ and is a direct generalization of the model previously obtained for the ML decoder in specific quantum codes~\cite{dennis2002topological,Wang:2002ph,Katzgraber:2009zz,Bombin:2012jk,Kubica:2018rab,Chubb:2021htd,Song:2021bud}.

Importantly, we note that the disordered stat-mech model $\calZ'$ exhibits an enlarged symmetry, which is a defining feature of the Nishimori line~\cite{nishimori1981internal}.
Specifically, the disorder average of $n-1$ copies of the partition function $\calZ'$, i.e.\ $\calZ_n$, exhibits a permutation symmetry $S_n$ over $n$ copies (as shown in Appendix~\ref{app:decoding_fidelity_error_configuration})~\cite{le1988location,gruzberg2001random}.
In our problem, the $n$-th moment $\calZ_n = \tr\rho_{Q\bfM}^n$ always has an $S_n$ symmetry which stems from the permutation symmetry over $n$ identical copies of the density matrices $\rho_{Q\bfM}$.

We further note that, for perfect measurements, $\bm{e}_t = \bm{a}_{\bm{v}_t} +\bar{\bm{a}}_{\bm{\kappa}_t}$ given by a stabilizer and a logical operator, and $\delta\bm{\epsilon}_t^1 = 0$. 
The stat-mech model is decoupled in the temporal direction, and the partition function $\calZ'$ reduces to
\begin{align}
    \calZ'(\bm{b}^1, \bm{\kappa}) &= \prod_t \sum_{\bm{v}_{t},\bm{\kappa}_t} \Pr\left(\bm{b}^1_t + \bar{\bm{a}}_{\bm{\kappa}_t} + \bm{a}_{\bm{v}_{t}}\right) \delta_{\sum_t \bm{\kappa}_t = \bm{\kappa}}.
\end{align}
The disordered stat-mech model for each time step is the same as the model in Ref.~\cite{Chubb:2021htd} derived for the ML decoder. 

\subsection{Duality between two expansions}\label{sec:stat-mech_duality}
Having derived stat-mech models based on the stabilizer expansion in Eq.~\eqref{eq:stabilizer_expansion} and error configuration expansion in Eq.~\eqref{eq:syndrome_expansion}, we here explicitly show that the two expansions are related by Fourier transformation.

Both expansions are possible because the set of stabilizers $g^{\bm{u}}$ and the set of syndrome projectors $\calP_{\bm{m}}$ both comprise operator bases for the same space, which is the set of density matrices that are block diagonal with respect to syndromes. In particular, we note that $\calP_{\bm{m}} = 2^{-I_0}\sum_{\bm{u}}(-1)^{\bm{m}\cdot\bm{u}}g^{\bm{u}}$ and $\ketbra{\bm{m}} = 2^{-I}\sum_{\bm{u}} (-1)^{\bm{m}\cdot\bm{u}}Z^{\bm{u}}$. Therefore, up to an overall constant,
\begin{equation}\label{eq:change_of_basis}
    \calP_{\bm{m}}\bigotimes_{t=1}^T \ketbra{\bm{m}'_t} = \sum_{\{\bm{u}_t\}}(-1)^{\sum_t\bm{\delta m}'_t\cdot\bm{u}_t}g^{\bm{u}_T}\prod_{t=1}^TZ_t^{\bm{u}_t-\bm{u}_{t-1}}
\end{equation}
where $\bm{\delta m}'_t = \bm{m}'_{t+1} - \bm{m}'_t$ with $\bm{m}'_{T+1} := \bm{m}$ and ${\bm{m}'_0 := \bm{0}}$. 
Eq.~\eqref{eq:change_of_basis} may also be inverted to express stabilizers in terms of syndromes.

By comparing Eq.~\eqref{eq:syndrome_expansion} and Eq.~\eqref{eq:stabilizer_expansion}, and applying the change of basis rule from Eq.~\eqref{eq:change_of_basis}, we arrive at a relationship between the coefficients in the two expansions:
\begin{equation}\label{eq:fourier_transform}
    \Pr(\bm{m}_T,\{\bm{m}'_t\}) = \sum_{\{\bm{u}_t\}}(-1)^{\sum_t\bm{\delta m}'_t\cdot\bm{u}_t}e^{-\calH(\{\bm{u}_t\})}.
\end{equation}
We observe that this may be viewed as a $\bbZ_2$ Fourier transform, and observe that it relates partition functions with quenched disorder in the error configuration expansion with Boltzmann weights in the stabilizer expansion.

As we saw in Sec.~\ref{sec:stat-mech_error_configuration}, $\Pr(\bm{m}_T,\{\bm{m}'_t\})$ may be written in terms of error configuration degrees of freedom. With this in mind, we can use Eq.~\eqref{eq:fourier_transform} to quickly obtain two dualities.

First, we observe that 
\begin{equation}
    \Pr(\bm{0},\{\bm{0}\}) = \sum_{\{\bm{u}_t\}}e^{-\calH(\{\bm{u}_t\})},
\end{equation}
which shows that the error configuration stat-mech model without disorder is dual to a single flavor of the stabilizer expansion stat-mech model.

Second, we observe that
\begin{equation}
    \sum_{\bm{m}_T,\{\bm{m}'_t\}}\Pr(\bm{m}_T,\{\bm{m}'_t\})^n = \sum_{\{\bm{u}_t^{\sfa}\}}e^{-\calH^{(n)}(\{\bm{u}^{\sfa}_t\})},
\end{equation}
since $\sum_{\bm{\delta m}} (-1)^{\sum_t\bm{\delta m}'_t\cdot\bm{u}_t}$ places a constraint on the $n$-th flavor of spins in terms of the other flavors, as we saw in Sec.~\ref{sec:stat-mech_stabilizer_quantities}. In this way, we see that the structure of the duality arises from a generalized form of the Plancherel theorem.

\subsection{Examples}

Having developed the mapping for generic stabilizer codes, we here consider the surface code, repetition code, and the XZZX code as specific examples and explicitly derive the stat-mech models to study their decoding transitions.
In what follows, we provide the summary of the results and leave the details in the appendices.

In Appendix~\ref{app:surface_code}, we consider the surface code subject to single-qubit Pauli errors and noisy measurements.
For Pauli-X (or Pauli-Z) error, the stat-mech model is a 3D $(n-1)$-flavor Ising model with nearest-neighbor couplings.
In Appendix~\ref{app:decoding_fidelity_error_configuration}, we verify that the model is the Kramers-Wannier dual of the 3D random plaquette Ising model in the limit $n\to 1$.
For Pauli-Y decoherence, the stat-mech model has two-body couplings in the temporal direction and four-body plaquette couplings in the spatial direction; the model is in the same universality as the $(2+1)$D Xu-Moore model~\cite{xu2004strong}.
We note that the $(2+1)$D dynamics of the surface code can be re-formulated as virtual time evolution generated by sequentially measuring the qubits in a 3D cluster state introduced by Raussendorf, Bravyi and Harrington (RBH)~\cite{raussendorf2005long}. 
In Appendix~\ref{app:RBH} we alternatively derive the stat-mech model for the $(2+1)$D surface code by considering the measurement dynamics of the noisy RBH state.

In Appendix~\ref{app:repetition_code}, we derive the stat-mech model for the $d$-dimensional repetition code subject to Pauli-X decoherence and noisy measurements. We show that the stat-mech model is a 2D $(n-1)$-flavor Ising model for the 1D repetition code, and the model is a 3D $(n-1)$-flavor $\bbZ_2$ gauge theory for the 2D repetition code. 

In Appendix~\ref{app:XZZX}, we consider the XZZX code subject to Pauli decoherence and noisy measurements. 
The resulting 3D $(n-1)$-flavor Ising model consists of two-body and four-body couplings, similar to the surface code. 
Notably, in the case of only Pauli-X (or Pauli-Z) decoherence, the model reduces to decoupled 2D Ising models.

\section{Discussion}\label{sec:discussion}
In this work, we study the dynamics of quantum memories under local Pauli decoherence channels and repeated noisy syndrome measurements. We develop a stat-mech model for non-linear functions of the density matrix, which allows evaluating information diagnostics, such as coherent information and relative entropy.

Our model is dual to the stat-mech model previously obtained for the optimal decoding algorithm. This duality can be understood by considering two different expansions of the same density matrix: the stabilizer expansion and the error configuration expansion. Writing the density matrix as a weighted sum of stabilizers leads to our model and writing the density matrix as a weighted sum of various errors applied to the initial state leads to the previously-obtained model.

Our dual model offers certain benefits. It exposes the impact of imperfect stabilizer measurements on information retention. The model consists of a layer of spins for each time step of the dynamics. After a single round of decoherence, quantities like the coherent information are determined by ordering in a single layer of spins. Perfect syndrome measurements ensure that successive layers of spins are completely decoupled from prior layers so that successive rounds of decoherence do not build on top of prior rounds. As the probability of readout noise increases, the layers of spins become coupled, making it possible for less information to be retained by the system over time. In the limit that readout noise is maximal, the layers are completely coupled together and it is as if no syndrome measurements occurred at all.

Our mapping also explains the necessity for repeating noisy measurements for successful error correction in the surface code even though the memory can in principle retain (maximum) coherent information even after $O(1)$ rounds of noisy measurements~\cite{dennis2002topological}.
This originates from the clear distinction between stat-mech models for $\rho_{Q\bfM}$ and $\rho_{\bfM}$. 
In particular, error correction requires identifying the error configuration only from the noisy measurement record and therefore concerns the stat-mech model associated with $\rho_{\bfM}$.
In the stabilizer expansion, this stat-mech model has a boundary magnetic field explicitly breaking the Ising symmetry and therefore cannot exhibit a paramagnetic phase for $O(1)$ rounds of noisy measurements.
When the noisy measurements are repeated for $O(L)$ times, the boundary magnetic field does not affect the ordering in the bulk of the (2+1)D stat-mech model.
Thus, the stat-mech model can be in the paramagnetic phase indicating successful error correction.
On the other hand, the coherent information is associated with the stat-mech model for $\rho_{Q\bfM}$, which has open boundary conditions on the final layer of spins and can undergo a paramagnetic-to-ferromagnetic transition for any finite number of rounds of noisy measurements.

Our results open several directions for further study. 
First, we note that in the dual picture developed in this work, i.e.\ the stat-mech model in the stabilizer expansion, codes with maximum threshold correspond to spin models that fail to order at any finite temperature. As such, one may be able to use insights from statistical mechanics to reverse engineer QEC codes that have maximal thresholds even in the presence of noisy measurements.

Second, there are certain QEC codes whose many redundancies or other special properties enable single-shot QEC~\cite{bombin2015single}, in which single rounds of noisy measurements can be used for decoding. Our dual stat-mech model may provide a useful platform for studying single-shot stabilizer QEC codes. Note that the stat-mech model describing moments of $\rho_\bfM$ after a single round of measurements would be a spin model with an external field, which normally induces ferromagnetic order. On the other hand, as we discussed in Sec.~\ref{sec:stat-mech}, local redundancies in measured stabilizers translate to gauge (or higher form) symmetries of the stat-mech model, and it is known that higher-form symmetries can be emergent even if the symmetry is explicitly broken microscopically~\cite{gaiotto2015generalized,mcgreevy2023generalized}. This may provide a useful perspective explaining the essential ingredients for successful single-shot decoding.

Third, although this work focuses only on stabilizer codes, it may be possible to generalize our stat-mech model to other related QEC codes, e.g. subsystem codes~\cite{poulin2005stabilizer} and dynamically generated codes~\cite{hastings2021dynamically}, subject to Pauli errors. 
In both cases, if starting from the maximally mixed state in the code space, the density matrix of the system at any time step can be expressed as a classical mixture of Pauli operators and, as such, there may exist natural extensions of our stat-mech model in each of these settings.

\emph{Note added:} After completing this manuscript, we became aware of a recent work \cite{niwa2024coherent}, which obtains results related to ours in Sec.~\ref{sec:setup_fidelity}.

\begin{acknowledgments}
We thank Ruihua Fan, Yaodong Li, Sagar Vijay, and Stephen Yan for helpful discussions. We thank Pavel Nosov for mentioning Ref.~\cite{fradkin1978gauge} to us. Y.B. thanks Ehud Altman, Ruihua Fan, and Ashvin Vishwanath for collaborating on previous works.
This material is based upon work supported by the
National Science Foundation Graduate Research Fellowship Program under Grant No. 2139319 (J.H.) and by the
Simons Collaboration on Ultra-Quantum Matter, which
is a grant from the Simons Foundation (651457, M.P.A.F., U.A., and J.H.).
M.P.A.F. is also supported by a Quantum Interactive Dynamics
grant from the William M. Keck Foundation.
Y.B. is supported in part by grant NSF PHY-2309135 and the Gordon and Betty Moore Foundation Grant No. GBMF7392 to the Kavli Institute for Theoretical Physics (KITP).
S.S. is supported by the Perimeter Institute for Theoretical Physics (PI) and the Natural Sciences and Engineering Research Council of Canada (NSERC).
Research at PI is supported in part by the Government of Canada through the Department of Innovation, Science and Economic Development Canada and by the Province of Ontario through the Ministry of Colleges and Universities. 
This research was also supported in part by the National Science Foundation under Grant No. NSF PHY-1748958 and NSF PHY-2309135, the Heising-Simons Foundation, and the Simons Foundation (216179, LB).

\end{acknowledgments}

\bibliography{refs}

\appendix

\section{Generalization to weak measurements}\label{app:weak_measurement}
In this work, we have focused on imperfect measurements where the imperfection is caused by a readout error that changes a measurement result. Weak measurements are another natural source of imperfections, where the unitary $U_{Q\to QM}$ fails to perfectly correlate the system and measurement ancilla, even before any readout errors occur. This interaction can be described by the same form of $U_{Q\to QM}$ in Eq.~\eqref{eq: U_Q,QM} but with $\calP_{0,1}$ modified such that
\begin{equation}
    \calP_{0,1} = \frac{1\pm \lambda g}{\sqrt{2(1+\lambda^2)}}
\end{equation}
where $0 \leq \lambda \leq 1$. When $\lambda = 1$ the operators $\calP_{0,1}$ are unchanged from the rest of the text, and the measurement becomes weaker the smaller $\lambda$ becomes.

As long as the initial state is in the code space and is subject only to Pauli errors and imperfect syndrome measurements, the state $\rho$ at any later time may be written as a weighted sum over stabilizers. Consequently, the state always commutes with $\calP_{0,1}$. Noting also that $\calP_0$ and $\calP_1$ commute with each other, and using the shorthand $U$ for $U_{Q\to Q\bfM}$, we can write
\begin{align}
    U \rho U^\dag = \rho[&\calP_0^2 \otimes \dyad{0}{0}_M +\calP_1^2 \otimes \dyad{1}{1}_M\nonumber\\ &+ \calP_0\calP_1\otimes(\dyad{0}{1}_M+\dyad{1}{0}_M)].
\end{align}
We note that
\begin{equation}
    \calP_{0,1}^2 = \frac{1}{2}\left(1 \pm \frac{2\lambda g}{1+\lambda^2}\right).
\end{equation}

It is useful to write $\dyad{0}{0}_M = (1+Z_M)/2$, $\dyad{1}{1}_M = (1-Z_M)/2$, and $\dyad{0}{1}_M+\dyad{1}{0}_M = X_M$, because the behavior of $Z_M$ and $X_M$ under readout errors is straightforward. In particular,
\begin{align}
    \calN_M[Z_M] &= (1-2q)Z_M\\
    \calN_M[X_M] &= X_M.
\end{align}
It is also useful to define weights $\calW_q = (1-2q)$ and $\calW_\lambda = 2\lambda/(1+\lambda^2)$. Then,
\begin{align}
    \calN_M[U\rho U^\dag] = \frac{1}{4}\rho[&(1+\calW_\lambda g)(1+\calW_qZ_M)\\
    &+ (1-\calW_\lambda g)(1-\calW_qZ_M)\\
    &+ \calP_0\calP_1 X_M].
\end{align}
In Sec.~\ref{sec:setup}, the $\calP_0\calP_1 X_M$ term was not present because $\calP_0\calP_1 = 0$ when $\lambda = 1$. As a result, it was not necessary to explicitly consider the impact of the actual projective measurement of the ancilla in the Pauli-Z basis. Here, it is necessary to consider this measurement, leading us to modify the imperfect measurement channel $\calM$:
\begin{multline}
\calM[\rho] = \left(\frac{1+Z_M}{2}\right)\calN_M[U\rho U^\dag]\left(\frac{1+Z_M}{2}\right)\\
+ \left(\frac{1-Z_M}{2}\right)\calN_M[U\rho U^\dag]\left(\frac{1-Z_M}{2}\right)
\end{multline}
which annihilates the $\calP_0\calP_1 X_M$ term. We conclude that
\begin{equation}
    \calM[\rho] = \frac{1}{2}\rho(1+\calW_\lambda\calW_q g Z_M).
\end{equation}
Analogous work in Sec.~\ref{sec:stat-mech_stabilizer_dynamics} yields the same result without the factor of $\calW_\lambda$. Since $\calW_\lambda$ is always positive, the combined weight $\calW_\lambda\calW_q$ is positive as long as $0 \leq q \leq \frac{1}{2}$. In Sec.~\ref{sec:stat-mech_stabilizer_dynamics} these weights and their logarithms are natural objects to consider, allowing us to interpret the impact of weak measurements as modifying
\begin{equation}
    \mu_q \to \mu_q - \log\frac{2\lambda}{1+\lambda^2}
\end{equation}
where $\mu_q = -\log(1-2q)$. We can also view this as modifying
\begin{equation}
    q \to q\left(\frac{2\lambda}{1+\lambda^2}\right) + \frac{1}{2}\frac{(1-\lambda)^2}{1+\lambda^2}
\end{equation}
which lies between $0$ and $\frac{1}{2}$ as long as $q$ does.

In summary, allowing weak measurements in our dynamics is equivalent to modifying the rate of readout errors. However, we note that the situation would be significantly more complicated in dynamics with coherent errors.

\section{Bound on the optimal decoding infidelity}\label{app:ML_infidelity}
Here, we provide a proof of the inequality
\begin{equation}
   \bar{\Delta} \leq H(\bm{\kappa}|X)
\end{equation}
where we use the shorthand notation $X:=(\bm{m}_T, \{\bm{m'}_t\})$, and
\begin{equation}
    \bar{\Delta} = \sum_X \Pr(X) (1 - \textrm{max}_{\bm{\kappa}} \Pr(\bm{\kappa} | X)).
\end{equation}
For fixed $X$, $H_{\infty}(\Pr(\bm{\kappa}|X)) = -\log \textrm{max}_{\bm{\kappa}}\Pr(\bm{\kappa}|X)$. It follows that
\begin{equation}
    \bar{\Delta} = \sum_{X}\Pr(X) (1- e^{-H_{\infty}(\Pr(\bm{\kappa}|X))}).
\end{equation}
Since $1-e^{-x}$ is concave, the linear approximation of the function at any point is an upper bound. Consequently,
\begin{equation}
    \bar{\Delta} \leq \sum_{X}\Pr(X) H_{\infty}(\Pr(\bm{\kappa}|X))
\end{equation}
and, using R\'enyi entropy inequalities,
\begin{equation}
    \bar{\Delta} \leq \sum_{X}\Pr(X) H(\Pr(\bm{\kappa}|X))
\end{equation}
where $H(\Pr(\bm{\kappa}|X)) = -\sum_{\bm{\kappa}} \Pr(\bm{\kappa}|X) \log \Pr(\bm{\kappa}|X)$. It follows that
\begin{align}
    \bar{\Delta} &\leq -\sum_{\bm{\kappa},X} \Pr(X)\Pr(\bm{\kappa}|X) \log \Pr(\bm{\kappa}|X)\\
    &= -\sum_{\bm{\kappa},X} \Pr(\bm{\kappa},X) \log \Pr(\bm{\kappa}|X)\\
    &= H(\bm{\kappa} | X).
\end{align}

\section{The $n \to 1$ limit of the stat-mech model in the error configuration expansion}\label{app:decoding_fidelity_error_configuration}
We here show that the disordered stat-mech model $\calZ'$ governs the information theoretical diagnostics in the limit $n \to 1$. We consider coherent information as an example.

We first take the error configuration expansion of the $n$-th moment $\tr\rho_{Q\bfM R}^n$
\begin{align}
    \tr\rho_{Q\bfM R}^n = &\sum_{\bm{b}^1,\bm{\epsilon}^1} \Pr(\bm{b}^1,\bm{\epsilon}^1) \calZ'(\bm{b}^1,\bm{\epsilon}^1,\bm{0})^{n-1}.
\end{align}
Here, the term in the error configuration expansion of $\rho_{Q\bfM R}$ contributes to $\tr\rho_{Q\bfM R}^n$ only if the cumulative error $\Sigma \bm{e}_T^\sfa = \Sigma \bm{b}_T^\sfa - \Sigma\bm{b}_T^1$ is given by a stabilizer, i.e.\ $\bm{\kappa} = [\Sigma\bm{e}_T^\sfa] = \bm{0}$.

Next, knowing that the partition function $\calZ'$ only depends on $(\bm{m}'_t, \bm{m}_T)$, we can express the $n$-th moment in terms of the probability of error strings belonging to different sectors, i.e.\ 
\begin{align}
    \tr\rho_{Q\bfM}^n &= \frac{1}{2^{K(n-1)}}\sum_{\bm{m}_T, \bm{m}'_t}\left(\sum_{\bm{\kappa}} \Pr(\bm{\kappa}, \bm{m}_T, \bm{m}'_t)\right)^n, \nonumber \\
    \tr\rho_{Q\bfM R}^n &= \sum_{\bm{m}_T, \bm{m}'_t}\sum_{\bm{\kappa}} \Pr(\bm{\kappa}, \bm{m}_T, \bm{m}'_t)^n,
\end{align}
where
\begin{align}
    \Pr(\bm{\kappa}, \bm{m}_T, \bm{m}'_t) & :=\calZ'(\bm{b}^1_0,\bm{\epsilon}^1_0,\bm{\kappa}).
\end{align}
Here, $\bm{b}^1_0,\bm{\epsilon}^1_0$ are a reference error configuration, and we have the relation $\calZ'(\bm{b}^1, \bm{\epsilon}^1, \bm{0}) = \Pr(\bm{\kappa}, \bm{m}_T, \bm{m}'_t)$ for $[\Sigma \bm{b}^1 - \Sigma \bm{b}_0^1] = \bm{\kappa}$.
This expression explicitly shows the enlarged $S_n$ permutation symmetry in the $n$-th moment $\calZ_n$.

In the limit $n \to 1$, the coherent information reduces to $I_c = K\log 2- H(\bm{\kappa}|\bm{m}_T, \bm{m}'_t)$ in agreement with Eq.~\eqref{eq:Ic_ml_decoder}, where $H(\bm{\kappa}|\bm{m}_T, \bm{m}'_t)$ is the Shannon entropy of the conditional distribution
\begin{align}
    \Pr(\bm{\kappa}|\bm{m}_T, \bm{m}'_t) = \frac{\calZ'(\bm{b}_0^1,\bm{\epsilon}_0^1,\bm{\kappa})}{\sum_{\bm{\kappa}}\calZ'(\bm{b}_0^1,\bm{\epsilon}_0^1,\bm{\kappa})}.
\end{align}
This calculation shows that the partition function $\calZ'$ determines the coherent information $I_c$.
Moreover, $\calZ'$ is proportional to the probability of each decoding sector and therefore is exactly the model obtained for the ML decoder~\cite{dennis2002topological}.

\section{Decoding transitions in the surface code}\label{app:surface_code}
The surface code~\cite{kitaev1997quantum,kitaev2003fault} is a promising quantum memory that can retain quantum information up to a finite threshold of decoherence and has been realized in leading experimental platforms~\cite{Semeghini:2021wls,Bluvstein:2021jsq,Satzinger:2021eqy,GoogleQuantumAI:2022fyn,Andersen:2022xmz,bluvstein2023logical}.
The information dynamics in the decohered surface code was previously studied in the context of optimal decoding algorithms~\cite{dennis2002topological}.
It was shown that the decoding transition based on a single round of measurements (with no readout errors) maps to a ferromagnetic transition in the 2D random bond Ising model, and the transition in the case of repeated imperfect measurements maps to a deconfinement transition in the 3D random plaquette Ising model (RPIM).
Here, we apply the stat-mech model developed in Sec.~\ref{sec:stat-mech} to the surface code and provide a dual perspective on the decoding transition in the case of repeated stabilizer measurements, generalizing the result in Ref.~\cite{fan2023diagnostics}.

As we explain below, the stat-mech model derived using the stabilizer expansion in Sec.~\ref{sec:stat-mech_stabilizer} exhibits a $\bbZ_2$ global symmetry and undergoes a ferromagnetic transition when tuning the decoherence rate, dual to the deconfinement transition in the RPIM. 
The decoherence rate tunes the temperature in the stat-mech model; the low-temperature ferromagnetic ordering corresponds to the information loss at a high decoherence rate.
Our dual picture makes it easier to understand situations with a maximum threshold, which manifest as stat-mech models exhibiting no finite-temperature ferromagnetic ordering.

In the rest of this section, we first introduce the surface code and develop the stat-mech model in the dual picture. 
We then consider the surface code subject to physical Pauli-X/Z errors under repeated imperfect measurements. We obtain a 3D $(n-1)$-flavor Ising model for the R\'enyi-$n$ quantities, which is dual to the previously known 3D RPIM in the limit $n \to 1$~\cite{dennis2002topological} (shown in Appendix~\ref{app:duality}).
Next, we consider the surface code subject to Pauli-Y errors, which is shown to have the maximum error threshold ($p_c = 0.5$) based on a single round of perfect measurements~\cite{tuckett2019tailoring}.
Using our stat-mech mapping, we show that unbiased measurement errors result in a sub-maximum threshold ($p_c < 0.5$).

\begin{table*}[t!]
    \centering
    \begin{tabular}{|c|c|c|c|}
         \hline\hline
         & \makecell{Coherent information \\ $I_c(R \rangle Q\bfM)$} & \makecell{Quantum relative entropy \\ $D_s(\rho_{Q\bfM}||\rho_{Q\bfM,s})$} & \makecell{KL divergence \\ $D_{\mathrm{KL},s}(\rho_{\bfM}||\rho_{\bfM,s})$}\\
         \hline
          Boundary field & $\times$ & $\times$ & $\checkmark$ \\
         \hline
         Finite threshold at $T = O(1)$ & $\checkmark$ & $\checkmark$ & $\times$ \\
         \hline
         Finite threshold at $T = O(L)$ & $\checkmark$ & $\checkmark$ & $\checkmark$ \\
         \hline\hline
    \end{tabular}
    \caption{Properties of the stat-mech model for information diagnostics in the decohered surface code under repeated stabilizer measurements.}
    \label{tab:diagnostics}
\end{table*}

\subsection{Surface code}
A surface code is a stabilizer code associated with any cellulation of a two-dimensional manifold.
We focus on the case of closed manifolds. Our results can be generalized to the planar surface code defined on an open manifold with suitable boundary conditions.
The cellulation defines a graph (2-complex) that consists of vertices (0-cells), edges (1-cells), and faces (2-cells).
The surface code consists of qubits on the edges $\ell$ of the graph and two types of stabilizers associated with vertices $v$ and faces $f$:
\begin{align}\label{eq:surface_code_stabilizers}
    A_v = \prod_{\ell \in \text{star}(v)} X_\ell, \quad B_f = \prod_{\ell \in \text{boundary}(f)} Z_\ell,
\end{align}
where star$(v)$ and boundary$(f)$ include all the edges emanated from the vertex $v$ and on the boundary of the face $f$, respectively.

The logical space of the surface code depends on the homology of the manifold.
Specifically, the stabilizer $B_f$ ($A_v$) is a product of Pauli operators along a boundary (co-boundary) on the graph.
The operators that commute with all the stabilizers are thus products of Pauli-Z(X) along cycles (co-cycles).
It follows that logical-Z(X) operators, which are not products of stabilizers, are given by the cycles (co-cycles) that are not boundaries (co-boundaries) and are thus determined by the homology (co-homology) group of the graph.
For a closed 2D manifold of genus $G$, the corresponding logical space can encode $K = 2G$ qubits.
See Ref.~\cite{bombin2007homological} for details.

\begin{figure}
    \centering
    \begin{tikzpicture}[]
    \pgftext{\includegraphics[width=0.49\linewidth]{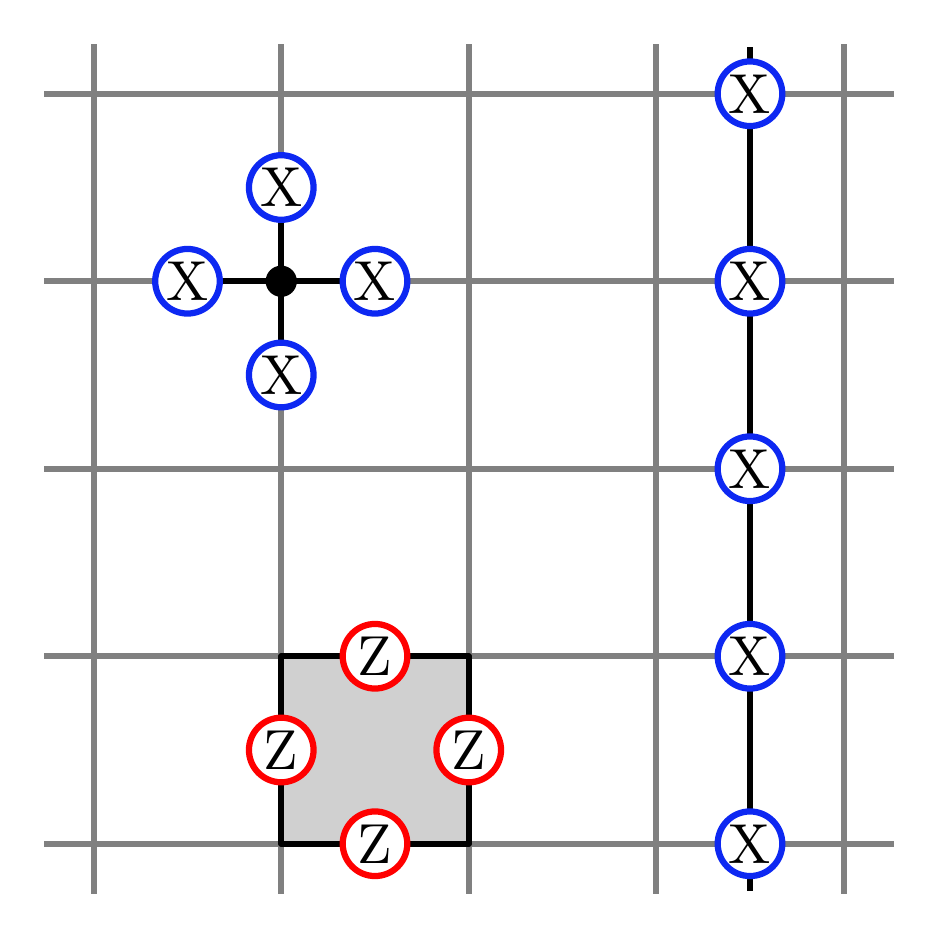}} at (0pt,0pt);
    \node at (-1.3, 1.3) {(a)};
    \node at (-1.3, -1.3) {(b)};
    \node at (1.05, 0.4) {(c)};
    \end{tikzpicture}
    \caption{Surface code on the square lattice with periodic boundary conditions. The code has stabilizers $A_v$ associated with vertices (a) and $B_f$ associated with plaquettes (b). The logical operator is a string operator along a non-contractible loop (c).}
    \label{fig:TC_definitions}
\end{figure}

A widely studied example of the surface code is the so-called toric code defined on a two-dimensional square lattice with periodic boundary conditions (topologically equivalent to a torus)~\cite{kitaev1997quantum,kitaev2003fault}.
The code space is four-fold degenerate and can encode two logical qubits.
The logical-Z(X) operators are string operators along two non-contractible loops on the direct (dual) lattice on the torus.
Examples of these check operators and logical operators are illustrated in Fig.~\ref{fig:TC_definitions}.
Our results in this section apply to general surface codes, but we use the toric code as a helpful example throughout.

We remark that the (2+1)D dynamics of the toric code under repeated syndrome measurements can be equivalently formulated as sequentially measuring the 3D Raussendorf-Bravyi-Harrington (RBH) state~\cite{raussendorf2005long}.
We detail this formulation and an alternative way to derive the stat-mech model in Appendix~\ref{app:RBH}.

\subsection{Stat-mech model of the surface code}
The stat-mech model for the surface code under repeated measurements involves Ising spins at each time step $t$ associated with stabilizer measurements of $A_v$ and $B_f$.
Thus, the Ising spins are defined on the vertices of the graph $\calG$ and the dual graph $\calG^*$ denoted by $\sigma_{r,t}$ and $\sigma_{\bar{r},t}$, respectively.
In the stabilizer expansion in Eq.~\eqref{eq:stabilizer_expansion_Zn}, the Boltzmann weight is given by an effective Hamiltonian
\begin{align}
    \calH = &\sum_{t,\ell} - J_z \sigma_{r,t}\sigma_{r',t} -J_x \sigma_{\bar{r},t}\sigma_{\bar{r'},t} - J_y \sigma_{r,t}\sigma_{r',t}\sigma_{\bar{r},t}\sigma_{\bar{r'},t} \nonumber \\
    &-\sum_{t,r} J_v \sigma_{r,t}\sigma_{r,t+1} - \sum_{t,\bar{r}}J_f \sigma_{\bar{r},t}\sigma_{\bar{r},t+1},
\end{align}
where $(r, r')$ is a pair of vertices connected by the edge $\ell$, $( \bar{r}, \bar{r'})$ is a pair of faces (vertices on the dual graph) connected by the dual edge $\ell^*$, and the couplings are given by Eq.~\eqref{eq:mu_defs} along with
\begin{align}
    J_v &= -\frac{1}{2}\log(1-2q_v), \\
    J_f &= -\frac{1}{2}\log(1-2q_f).
\end{align}
Here, $q_v$ and $q_f$ denote the readout error rates for measuring $A_v$ and $B_f$, respectively.

We can now write down the $n$-th moment as the partition function of a stat-mech model of $n-1$ flavors of Ising spins, $\tr\rho^n_{Q\bfM} = \calZ_n$.
The symmetry of the stat-mech model is determined by the redundancies of the check operators.
Here, the product of all star operators as well as the product of all plaquette operators is the identity, leading to the redundancy group $\calR \cong \bbZ_2\times \bbZ_2$.
Hence, the stat-mech model for the $n$-th moment exhibits a global symmetry of $\mathbb{G}^{(n)} = (\calR^n\rtimes S_n)/\calR$.
When increasing the decoherence rate above the threshold, we expect a transition from a symmetric paramagnetic phase to a ferromagnetic phase that breaks the $\mathbb{G}^{(n)}$ symmetry.
In what follows, we formulate the information diagnostics as observables in the stat-mech model and show they indeed detect the transition.

\emph{Coherent information.---} Using Eq.~\eqref{eq:stabilizer_expansion_coherent_info}, the $n$-th moment $\tr\rho_{RQ\mathbf{M}}^n$ can be expressed as 
\begin{align}
    \tr \rho_{RQ\mathbf{M}}^n = \frac{1}{2^{2(n-1)}}\sum_{\{\bm{\kappa}^\sfa\}}\calZ_n(\{\bm{\kappa}^\sfa\}),
\end{align}
where $\bm{\kappa}^\sfa$ for $\sfa=1,\dots,n-1$ is a $4$-component binary vector; the first (last) two components label the insertion of two-dimensional non-contractible defects in $xt$ and $yt$ planes associated with logical-X(Z) operators, respectively.

We thus can write the $n$-th coherent information as 
\begin{align}
    I_c^{(n)} = \frac{1}{n-1} \log \left(\sum_{\{\bm{\kappa}^\sfa\}}\frac{\calZ_n(\{\bm{\kappa}^\sfa\})}{\calZ_n}\right) - 2 \log 2.
\end{align}
We note that $-\log (\calZ_n(\{\bm{\kappa}^\sfa\})/\calZ_n) = \Delta F_n(\{\bm{\kappa}^\sfa\})$ is the excess free energy of inserting the defects denoted by $\bm{\kappa}^\sfa$.
In the paramagnetic phase, the defect costs zero free energy in the thermodynamic limit, so all the $2^{4(n-1)}$ terms in the summation are unity, thus, $I_c^{(n)} = 2\log 2$.
In contrast, in the ferromagnetic phase of $\sigma^\sfa$ associated with Pauli-Z strings (assuming only Pauli-X errors and no Pauli-Z errors), the defect associated with logical-Z operators costs a free energy $O(LT)$ that increases with the system size. In this case, only $2^{2(n-1)}$ terms in the summation are unity, and we have $I_c^{(n)} = 0$.
Hence, the decoding transition in the R\'enyi-$n$ coherent information can be understood as the ordering transition in the 3D stat-mech model.

\emph{Relative entropy.---} Here, we find the R\'enyi-$n$ relative entropy in the $(n-1)$-flavor Ising model. We consider the relative entropy between $\rho_{Q\bfM}$ and $\rho_{Q\bfM,m}$ where the latter state is evolved from the excited state $\rho_{m} = w_{m}(\gamma)\rho_0w_{m}^\dagger(\gamma)$ with $w_{m}(\gamma) = \prod_{\ell\in\gamma} X_\ell$ creating $m$ anyons at the endpoints $r_1$ and $r_2$ of $\gamma$.

Then, according to Eq.~\eqref{eq:stabilizer_expansion_relative_entropy}, the $n$-th R\'enyi relative entropy maps to the spin-spin correlation function on the bottom boundary of the stat-mech model
\begin{align}
    D^{(n)}_m = -\frac{1}{n-1} \log \left\langle \sigma^1_{0,r_1} \sigma^1_{0,r_2} \right\rangle.
\end{align}
Hence, the relative entropy detects the ferromagnetic transition in the $(n-1)$-flavor Ising model; it diverges in the paramagnetic phase and takes a finite value in the ferromagnetic phase.

Similarly, the KL divergence $D_{\text{KL},m}^{(n)}$ maps to the correlation function in the stat-mech model with a boundary field.
When the number of time steps $T = O(L)$, the model can undergo a ferromagnetic phase transition in the bulk, which is detected by the KL divergence.

\subsection{Threshold of Pauli-X/Z error} 
The first example we consider is the surface code subject to Pauli-Z (or -X) errors on the physical qubits, i.e.\ $p_x = p_y = 0$ (or $p_y = p_z = 0$).
We focus on the case of Pauli-Z errors. The analysis for Pauli-X is similar.

The Ising spins on the direct and dual graph are decoupled in the stat-mech model, and the only non-vanishing couplings are $J_z = -(1/2)\log(1-2p_z)$, $J_v$, and $J_f$. 
We now consider the $n$-th moment $\tr\rho_{Q\bfM}^n$ and write the effective Hamiltonian of the stat-mech model.

The stat-mech model for the spins on the dual graph $\calG^*$ is decoupled one-dimensional Ising models with ferromagnetic couplings between neighboring sites in the temporal direction, which are always disordered for $q_f < 0.5$.
This is consistent with the fact that Pauli-Z errors do not flip the eigenvalues of $B_f$ stabilizers, and the classical information encoded in the logical Pauli-Z is always retained.

The stat-mech model for the spins on the direct graph $\calG$ is a 3D multiple-flavor Ising model
\begin{align}
    \calH^{(n)}_Z = \sum_{\ell,t}  &-J_z \sum_{\sfa = 1}^{n-1}\sigma_{r,t}^\sfa \sigma_{r',t}^\sfa - J_z\prod_{\sfa = 1}^{n-1}\sigma_{r,t}^\sfa \sigma_{r',t}^\sfa\nonumber \\
    + \sum_{r,t} &-J_v \sum_{\sfa=1}^{n-1}\sigma_{r,t}^\sfa\sigma_{r,t+1}^\sfa - J_v \prod_{\sfa=1}^{n-1}\sigma_{r,t}^\sfa \sigma_{r,t+1}^\sfa, \label{eq:HZn}
\end{align}
where $\sfa = 1,2,\dots, n-1$ label the Ising spins in the $\sfa$-th replica copy.
The model is dual to the RPIM in the limit $n \to 1$ as shown in Appendix~\ref{app:duality}.

The model undergoes a paramagnetic-to-ferromagnetic phase transition when increasing the error rate. In the case of $n = 2$, the stat-mech model reduces to a 3D Ising model with couplings $2J_z$ and $2J_v$. In the isotropic limit $p = q$ (i.e.\ $J_z = J_v$), the transition occurs at $2\mu_{z,c}^{(2)} = 0.221$~\cite{barber1985finite,ferrenberg1991critical}, corresponding to $p_c^{(2)} = q_c^{(2)} = 0.099$.
In the limit $n \to \infty$, the inter-replica interaction is negligible, giving rise to a 3D Ising model with couplings $J_z$ and $J_v$. In the isotropic limit $p = q$, the transition occurs at $p_c^{(\infty)} = q_c^{(\infty)} = 0.179$.

\subsection{Threshold of Pauli-Y error}
The next example we consider is the surface code with Pauli-Y errors ($p_x = p_z = 0$). 
The Ising spins on the direct and dual graph are coupled in the stat-mech model.
The couplings are given by $J_x = J_z = 0$ and $J_y = -(1/2)\log(1-2p_y)$.
We remark that, in this case, the stat-mech model exhibits subsystem symmetries in each flavor.
In what follows, we analyze the threshold for different types of measurement errors.

\subsubsection{Perfect measurement}
We start by considering the case with perfect measurements. Here, the 3D multi-flavor Ising model consists of decoupled 2D multi-flavor Ising models in the temporal direction. The 2D model has a Hamiltonian
\begin{align}
    \calH^{(n)}_{Y,0} = -J_y\sum_{\ell} \left(\sum_{\sfa = 1}^{n-1}\sigma^\sfa_r \sigma^\sfa_{r'} \sigma^\sfa_{\bar r} \sigma^\sfa_{\bar r'} + \prod_{\sfa = 1}^{n-1} \sigma^\sfa_r \sigma^\sfa_{r'} \sigma^\sfa_{\bar r} \sigma^\sfa_{\bar r'}\right).
\end{align}
The model is a plaquette Ising model in 2D, which exhibits no finite temperature order for any replica index $n$. The absence of finite temperature order indicates a maximum threshold $p_c^{(n)} = 0.5$. This is consistent with $p_c = 0.5$ in the replica limit~\cite{tuckett2019tailoring}.

We remark that the maximum threshold in the case of Pauli-Y errors is relevant for realizing the surface code in experimental platforms with biased noise, i.e.\ when Pauli errors along a specific direction dominate.
One can then tailor the surface code such that the biased noise is in the Pauli-Y direction and allow the code to achieve a high threshold of encoding quantum information~\cite{tuckett2019tailoring}.
With the same motivation, the XZZX code, related to the surface code by local rotations, has been proposed to encode information up to a maximum threshold of Pauli-X, Y, or Z errors~\cite{bonilla2021xzzx}.

\subsubsection{Noisy measurement}
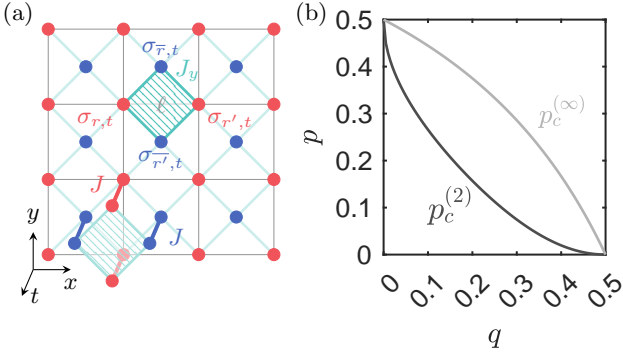
\begin{figure}[t!]
    \centering
    \begin{tikzpicture}
        \definecolor{myred}{RGB}{240,83,90};
    \definecolor{myblue}{RGB}{73,103,189};
    \definecolor{myturquoise}{RGB}{83,195,189};
    \small
    \foreach \y in {0,1,...,3}{
        \draw[black!40] (-0.,\y) -- (3.,\y);
    }
    \foreach \x in {0,1,...,3}{
        \draw[black!40] (\x,-0.) -- (\x,3.);
    }
    
    \foreach \x in {0,1,2}{
    \foreach \y in {0,1,2}{
    \draw[myturquoise!30,line width=1] (\x,\y) -- (\x+0.5,\y+0.5);
    }
    }
    \foreach \x in {1,2,3}{
    \foreach \y in {0,1,2}{
    \draw[myturquoise!30,line width=1] (\x,\y) -- (\x-0.5,\y+0.5);
    }
    }
    \foreach \x in {0,1,2}{
    \foreach \y in {1,2,3}{
    \draw[myturquoise!30,line width=1] (\x,\y) -- (\x+0.5,\y-0.5);
    }
    }
    \foreach \x in {1,2,3}{
    \foreach \y in {1,2,3}{
    \draw[myturquoise!30,line width=1] (\x,\y) -- (\x-0.5,\y-0.5);
    }
    }
    \draw[black!5] (1,2) -- (2,2);
    \draw[pattern=north west lines, pattern color=myturquoise, draw=myturquoise, line width=1, rotate around={-45:(1,2)}] (1,2) rectangle (1.707106781, 2.707106781);

    \node[myred, below left] at (1,2) {$\sigma_{r,t}$};
    \node[myred, below right] at (2,2) {$\sigma_{r',t}$};
    \node[myblue, above] at (1.5,2.5) {$\sigma_{\bar{r},t}$};
    \node[myblue, below] at (1.5,1.5) {$\sigma_{\bar{r'},t}$};
    \node[black!40] at (1.5,2) {$\ell$};
    \node[myturquoise] at (1.85,2.45) {$J_y$};
    \node[myred] at (0.65,0.94) {$J$};
    \node[myblue] at (1.7,0.23) {$J$};
           
    \foreach \y in {0.5,1.5,...,2.5}{
    \foreach \x in {0.5,1.5,...,2.5}{
    \filldraw[myblue] (\x,\y) circle (0.08);
    }
    }
    \foreach \y in {0,1,...,3}{
    \foreach \x in {0,1,...,3}{
        \filldraw[myred] (\x,\y) circle (0.08);
    }
    }
    
    \draw[myblue, line width=1.5] (1.5,0.5) -- (1.35,0.15);
    \draw[myblue, line width=1.5] (0.5,0.5) -- (0.35,0.15);
    \draw[myred,line width=1.5] (1,1) -- (0.85,0.65);
    \draw[myred, line width=1.5] (1,0) -- (0.85,-0.35);
    \fill[white, semitransparent,rotate around={-45:(0.35,0.15)}] (0.35,0.15) rectangle (1.057106781, 0.857106781);
    \draw[pattern=north west lines, pattern color=myturquoise!50, draw=myturquoise!50,line width=1,rotate around={-45:(0.35,0.15)}] (0.35,0.15) rectangle (1.057106781, 0.857106781);
    \filldraw[myblue] (0.35,0.15) circle (0.08);
    \filldraw[myblue] (1.35,0.15) circle (0.08);
    \filldraw[myred] (0.85,0.65) circle (0.08);
    \filldraw[myred] (0.85,-0.35) circle (0.08);
    
    \draw[->,>=stealth] (-0.2,-0.2) -- (-0.2,0.3) node[above]{$y$};
    \draw[->,>=stealth] (-0.2,-0.2) -- (0.3,-0.2) node[below]{$x$};
    \draw[->,>=stealth] (-0.2,-0.2) -- (-0.35,-0.55) node[right]{$t$};
    \node[inner sep=0pt] at (5.5,1.0) {\includegraphics[width=0.5\linewidth]{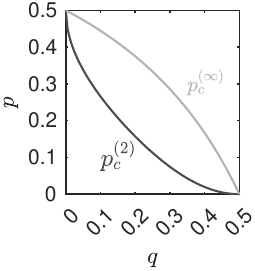}};
    \node at (-0.4,3.2) {(a)};
    \node at (3.5,3.2) {(b)};
    \end{tikzpicture}
    \caption{Critical error rates $(p_c^{(n)}, q_c^{(n)})$ in the $(n-1)$-flavor Ising model for the toric code with Pauli-Y errors and syndrome measurement errors. The physical and measurement error rates are denoted by $p$ and $q$, respectively. The critical points are determined analytically for $n = 2$ (dark) and $n \to \infty$ (light) based on the self-duality of the model.}
    \label{fig:critical_point_Pauli-Y}
\end{figure}

We now consider the surface code subject to Pauli-Y errors and noisy measurements with unbiased readout errors. 
In this case, in addition to the four-body Ising couplings among spins at the same time step, Ising spins on consecutive time steps are also coupled by two-body Ising terms. 
The result is an Ising spin model on a cubic lattice (tilted from the original lattice by $\pi/4$) that contains $2L^2$ spins in each one of the $T$ time steps (illustrated in Fig.~\ref{fig:critical_point_Pauli-Y}(a)).
The $n$-th moment maps the partition function with the Hamiltonian
\begin{align}
    \calH_Y^{(n)} = \quad &\nonumber \\
    -J_y\sum_{\ell,t}& \sum_{\sfa = 1}^{n-1}\sigma^\sfa_{r,t} \sigma^\sfa_{r',t} \sigma^\sfa_{\bar{r},t} \sigma^\sfa_{\bar{r'},t} + \prod_{\sfa = 1}^{n-1} \sigma^\sfa_{r,t} \sigma^\sfa_{r',t} \sigma^\sfa_{\bar{r},t} \sigma^\sfa_{\bar{r'},t} \nonumber \\
    -J \sum_{r, t} & \sum_{\sfa=1}^{n-1}\sigma_{r,t}^\sfa\sigma_{r,t+1}^\sfa + \prod_{\sfa=1}^{n-1}\sigma_{r,t}^\sfa \sigma_{r,t+1}^\sfa\nonumber \\
    -J \sum_{\bar{r}, t} & \sum_{\sfa=1}^{n-1}\sigma_{\bar{r},t}^\sfa\sigma_{\bar{r},t+1}^\sfa + \prod_{\sfa=1}^{n-1}\sigma_{\bar{r},t}^\sfa \sigma_{\bar{r},t+1}^\sfa.
\end{align}
where $J := J_v = J_f = -(1/2)\log(1-2q)$, and $q$ is the unbiased readout error rate.

The phase boundary of the model can be analytically determined for certain replica indices.
For $n = 2$, the model is self-dual~\cite{xu2004strong}. Assuming only two phases of the model, a ferromagnetic and a paramagnetic phase, the transition occurs at the self-dual line satisfying $\sinh(4J_y)\sinh(4J) = 1$, i.e.\
\begin{align}
    p_c^{(2)} = \frac{1}{2}\left( 1 - \sqrt{\frac{1 - \left(1-2q_c^{(2)}\right)^2}{1 + \left(1-2q_c^{(2)}\right)^2}}\right).
\end{align}
In the limit $n \to \infty$, neglecting the inter-replica interaction, the self-dual line is given by $\sinh(2J_y)\sinh(2J) = 1$, i.e.\
\begin{align}
    p_c^{(\infty)} = 1-\frac{1}{2 - 2q_c^{(\infty)}}.
\end{align}
The critical points for $n = 2$ and $n \to \infty$ are illustrated in Fig.~\ref{fig:critical_point_Pauli-Y}(b).
Away from the limit of perfect measurement $q = 0$, the stat-mech model exhibits a finite temperature ferromagnetic phase, indicating a threshold that is less than the maximum value, i.e.\ $p_c < 0.5$ for $q > 0$.
We remark that for $n = 2$ and the limit of $n \to \infty$, the 3D stat-mech model is in the universality of the (2+1)D quantum Xu-Moore model~\cite{xu2004strong}.

\section{Duality between random plaquette Ising model and multi-flavor Ising model in 3D}\label{app:duality}
In this section, we explicitly derive the duality between the 3D multi-flavor Ising model in Sec.~\ref{sec:stat-mech} and the 3D random plaquette Ising model (RPIM) derived for error correction in the toric code with Pauli-X/Z errors under repeated erroneous syndrome measurements~\cite{dennis2002topological}.

The 3D RPIM derived in Ref.~\cite{dennis2002topological} is of the form
\begin{align}
    H_\text{RPIM} = -\sum_{\sfp\in\square_{xy}} \tilde{J}_q\eta_\sfp u_\sfp-\sum_{\sfp \in \square_{xt}\cup \square_{yt}} \tilde{J}_p \zeta_\sfp u_\sfp,
\end{align}
where $\tilde{J}_p = \log \sqrt{(1-p)/p}$, $\tilde{J}_q = \log \sqrt{(1-q)/q}$, $u_\sfp = \prod_{\bfr \in \partial \sfp}\tau_{\bfr}$, and $\eta_\sfp, \zeta_\sfp = \pm 1$ are random variables that take~$-1$ with probability $p$ and $q$, respectively.
Importantly, the Hamiltonian has a constraint for $\sfp \in \square_{xy}$ on the bottom layer; the product of four spins around a plaquette is forced to be $+1$.
In the language of $\mathbb{Z}_2$ gauge theory, the flux line does not penetrate the bottom boundary.
The spins on the top boundary are not subject to hard constraints, and the flux line can terminate at the top boundary.
The RPIM model has periodic boundary conditions in the $x$ and $y$ directions.
Here, for the toric code on a 2D square lattice of size $L \times L$ under $T$ rounds of repeated syndrome measurements, the RPIM model is defined on a 3D slab of size $L \times L \times T$ and consists of spins on the edges ($3L^2T$ spins in total).

Before deriving the duality, we first rewrite the partition function of the 3D RPIM in terms of the flux lines in the 3D classical $\bbZ_2$ gauge theory.
Any spin configuration can be obtained by flipping a subset of spins to the $-1$ state from the configuration of every spin in the $+1$ state.
The spin-flip creates the violation of plaquette variables along a closed loop $\calC$ on the dual lattice, i.e.\ the $\pi$-flux line.
This allows writing $\calZ_\text{RPIM}$ as
\begin{align}
    \calZ_{\text{RPIM}} = \; &e^{\tilde{J}_pN_t(1-2p)+\tilde{J}_qN_{xy}(1-2q)} \nonumber \\
    &\sum_{\calC} \prod_{\sfp \in \calC_{xy}} e^{-2\tilde{J}_p\eta_\sfp} \prod_{\sfp \in \calC_t} e^{-2\tilde{J}_q\eta_\sfp},
\end{align}
where $N_{xy}$ ($N_t$) is number of plaquettes in the $xy$ plane ($xt$ and $yt$ plane), and $\calC_{xy}$ ($\calC_t$) is the collection of plaquettes in the $xy$ plane ($xt$ and $yt$ plane) on the loop $\calC$.

The RPIM partition function is dual to a nonlocal correlation function in a 3D Ising model with spins defined at the center of cubes in the original lattice,
\begin{align}
    \calZ_{\text{Ising}} =\;& \sum_{\{s_i\}} e^{J_p\sum_{\langle i, j\rangle_{xy}}s_i s_j + J_q\sum_{\langle i,j \rangle_t} s_i s_j} \prod_{\eta_\sfp = -1}s_i s_j \nonumber \\
    =\;& (\cosh J_p)^{N_t(1-p)} (\sinh J_p)^{N_t p}\nonumber \\
    & (\cosh J_q)^{N_{xy}(1-q)} (\sinh J_q)^{N_{xy} q} \nonumber \\
    & 2^N\sum_{\calC} \prod_{\sfp \in \calC_{xy}}(\tanh J_p)^{\eta_\sfp} \prod_{\sfp \in \calC_{t}}(\tanh J_q)^{\eta_\sfp},
\end{align}
where the link $\langle i,j \rangle$ is dual to the plaquette $\sfp$, the subscript $xy$ ($t$) denotes the plaquette in the $xy$ plane ($xt$ and $yt$ plane), and $N$ is the total number of spins.
The two partition functions are identical up to a non-singular prefactor provided that $\tanh J_p = e^{-2\tilde{J}_p}$ and $\tanh J_q = e^{-2\tilde{J}_q}$.
Such a duality is obtained in Ref.~\cite{fradkin1978gauge}.

Lastly, we show that the average $(n-1)$-th moment of the RPIM partition function, and equivalently that of $\calZ_{\text{Ising}}$, maps to the partition function of the $(n-1)$-flavor Ising model derived in Sec.~\ref{sec:stat-mech}.
Here, we obtain the effective Hamiltonian by averaging over random plaquette variables $\eta_\sfp$ and $\zeta_\sfp$
\begin{align}
    \overline{\calZ_{\text{Ising}}^{n-1}} = \sum_{\{s_i\}} \;&\; e^{\sum_{\sfa = 1}^{n-1}J_p\sum_{\langle i, j\rangle_{xy}}s_i s_j + J_q\sum_{\langle i,j \rangle_t} s_i s_j} \nonumber \\
    &\;\prod_{\langle i,j \rangle_{xy}} \left(1- p + p \prod_{\sfa = 1}^{n-1} s_i^\sfa s_j^\sfa\right) \nonumber \\
    &\;\prod_{\langle i,j \rangle_{t}} \left(1- q + q \prod_{\sfa = 1}^{n-1} s_i^\sfa s_j^\sfa\right) \nonumber \\
    = \sum_{\{s_i\}} \;&\; e^{-H_Z^{(n)}}.
\end{align}
The effective Hamiltonian is exactly the one obtained in Eq.~\eqref{eq:HZn}.

We here focus on the duality between the two models in the bulk.
In Appendix~\ref{app:boundary_condition}, we comment on the boundary conditions in the RPIM and its dual.

\section{Stat-mech model in the limit of perfect measurements}\label{app:boundary_condition}
In this section, we consider the limit of perfect measurement in the multi-flavor Ising model obtained in the dual picture. 
We show that perfect syndrome measurements lead to a stat-mech model with open boundary conditions at the top of the slab. 
The resulting stat-mech model exhibits a finite-temperature ordering transition giving rise to a finite error threshold regardless of how many rounds of syndrome measurements are performed.

The couplings in the multi-flavor Ising model in the temporal direction are determined by the rate of measurement errors, $J_q = -(1/2)\log (1 - 2q)$; the coupling $J_q$ is vanishing if $q = 0$. 
Thus, if the final round of measurement is perfect, the model has an open boundary condition at the top and, therefore, exhibits a transition at a finite threshold regardless of the thickness $T$.

The existence of a finite decoding threshold can be also understood in the original picture of the RPIM. 
Here, the boundary magnetic field in the dual picture translates to the boundary condition for magnetic $\pi$-flux lines. 
The $\pi$-flux line is allowed to terminate at the top boundary when the boundary field is nonzero. 
We note that the multi-flavor Ising model always has an open boundary condition at the bottom, i.e.\ $\pi$-flux line cannot terminate at the bottom. 
In this case, if the slab has a finite thickness $T$, there is no transition when tuning the error rate because the 2D classical $\bbZ_2$ gauge theory is always confined in the effective 2D slab.
In contrast, when the fluxes are \emph{not} allowed to terminate at both the top and the bottom, the gauge theory has a special boundary condition, which allows the transition to happen.

\section{Toric code with repeated syndrome measurements as 3D RBH}\label{app:RBH}
In this section, we briefly explain how the 3D cluster state  
of Raussendorf, Bravyi and Harrington (RBH)~\cite{raussendorf2005long} can be used to reason about the quantum information content of the 2D toric code under repeated noisy measurements in the presence of noise. Although the discussion in this section is focused on the 2D toric code, it readily generalizes to any CSS code, by replacing the RBH state with the corresponding foliated cluster state~\cite{bolt2016foliated}.
\subsection{The RBH state}
\begin{figure}
    \centering
    \includegraphics{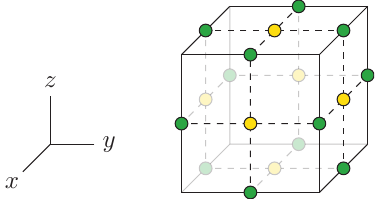} 
    \caption{The unit-cell of the 3D cubic lattice $\Lambda$ used to define the RBH state. The qubits are placed on the edges (green) and the faces (yellow) of $\Lambda$. The RBH state corresponds to a graph state, with the underlying graph specified by the dashed lines.}
    \label{fig:rbh_unitcell}
\end{figure}
Consider the 3D cubic lattice $\Lambda$ with qubits placed on edges and faces (Fig.~\ref{fig:rbh_unitcell}). Let $E$ and $F$ denote the set of edges and faces of $\Lambda$. Let $P_e$ ($P_f$) denote the Pauli operator $P$ acting on the qubit which is placed on $e\in E$ ($f \in F$). For each edge, define $g_e=X_e \prod_{f: e\in \partial f}Z_f$, where $\partial f$ denotes the boundary of $f$. Similarly, for each face $f\in F$ define $g_f=X_f \prod_{e \in \partial f} Z_e$. The RBH state is the state stabilized by $g_e$ and $g_f$ for all $e\in E$ and $f \in F$. Equivalently, the RBH state can viewed as the graph state~\cite{van2004graphical} corresponding to the graph obtained by connecting each face qubit to the four edge qubits on the face boundary (the dashed line in Fig.~\ref{fig:rbh_unitcell}).  As such, the RBH sate $\ket{\psi_\text{RBH}}$ can be written as,
\begin{align}
    &\ket{\psi_\text{RBH}}=U_\Lambda\ket{+}^{\otimes N}\\
    &U_\Lambda=\prod_{f\in F} \prod_{e\in\partial f} \text{CZ}_{e,f} \label{eq:ulambda}
\end{align}
where $\text{CZ}_{i,j}$ stands for the controlled $Z$ gate between qubits $i$ and $j$. Thus one could prepare the RBH state by initializing every qubit in the $\ket{+}$ state and applying CZ gates along the dashed lines in Fig.~\ref{fig:rbh_unitcell}.

\begin{figure}
 \begin{subfigure}{0.49\columnwidth}
     \includegraphics[width=0.75\textwidth]{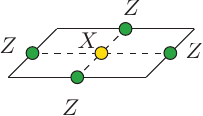}
     \caption{}
     \label{fig:rbh_plaquette}
 \end{subfigure}
 \hfill
 \begin{subfigure}{0.49\columnwidth}
     \includegraphics[width=\textwidth]{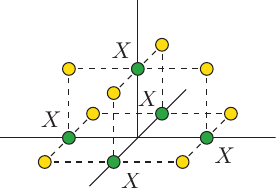}
     \caption{}
     \label{fig:rbh_star}
 \end{subfigure}
 
 \medskip
 \begin{subfigure}{0.49\columnwidth}
     \includegraphics[width=0.95\textwidth]{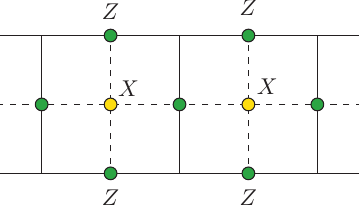}
     \caption{}
     \label{fig:rbh_zlogical}
 \end{subfigure}
 \hfill
 \begin{subfigure}{0.49\columnwidth}
     \includegraphics[width=0.95\textwidth]{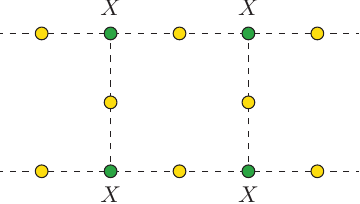}
     \caption{}
     \label{fig:rbh_xlogical}
 \end{subfigure}

 \caption{a) The $g_f$ stabilizer on the boundary is equal to the toric code plaquette operator on the boundary times $X_f$. b) The product of four $g_e$ stabilizers around a vertex on the boundary is equal to the toric code star operator on the boundary times $X_e$ in the bulk. c) The product of all $g_f$ stabilizers that lie on a vertical plane is equal to the product of the toric code $Z$ logical operators on the top and bottom boundary times a product of some $X_f$ in the bulk. d) The product of $g_e$ operators where $e$ cuts through a fixed veritcal plane is equal to the product of the toric code $X$ logical operators on the top and bottom boundary times product of some $X_f$ in the bulk.  }
 \label{Label}

\end{figure}

\subsection{The RBH state and the toric code}\label{apx_sub:rbh_Toric_bell}
Consider the RBH state on the $L_x\times L_y\times L_z$ cubic lattice with periodic boundary conditions along the $x$ and $y$ axes and open boundary condition along the $z$ axis. If one measures all qubits except the edge qubits on the top and bottom boundaries in the Pauli-X basis, the resulting state on the unmeasured qubits consists of toric code eigenstates on the boundaries, which are entangled to make a pair of logical Bell pairs~\cite{raussendorf2005long}. To see this, note that a face stabilizer $g_f$ on a boundary is just the toric code plaquette operator times $X_f$ (Fig.~\ref{fig:rbh_plaquette}). Measuring the face qubit in the Pauli-X basis thus ensures that the unmeasured qubits are an eigenstate of the corresponding toric code plaquette operator, with the eigenvalue determined by the measurement outcome of $X_f$. Moreover, if one multiplies all the five edge stabilizers $g_e$ that are stemming out of a vertex at a boundary, one obtains the star operator of the toric code on the boundary times $X_e$ in the bulk (Fig.~\ref{fig:rbh_star}). Thus, measuring the bulk qubits in the Pauli-X basis projects the boundary edge qubits into an eigenstate of the toric code star operator. Lastly, multiplying all the face stabilizers $g_f$ where $f$ lies in a fixed $yz$ plane (or in a fixed $xz$ plane) gives the product of the $Z$ logical operators of the toric codes on the top and bottom boundaries times a product of $X_f$ in the bulk (Fig.~\ref{fig:rbh_zlogical}). Similarly, the product of all edge stabilizers $g_e$ where $e$ cuts through a fixed $yz$ plane (or a fixed $xz$ plane) gives the product of the $X$ logicals times a product of $X_f$ in the bulk (Fig.~\ref{fig:rbh_xlogical}). Therefore, measuring bulk qubits in the $X$ basis projects the boundary toric codes into logical Bell pairs, with the particular Bell pairs determined by the signs of the bulk Pauli-X measurements.

\subsection{Using the RBH state to teleport toric code eigenstates}
Just like the 1D cluster state can be used to teleport a single qubit~\cite{raussendorf2000quantum,raussendorf2009measurement}, the 3D cluster state, i.e.\ the RBH state, can be used to teleport a 2D toric code state along the 3rd axis. Consider the following setup. Imagine having qubits on the edges and faces of a $L_x\times L_y \times 1$ slab of a cubic lattice $\Lambda$  with periodic boundary conditions along $x$ and $y$ directions, and open boundaries at the $z=0$ and $z=1$ planes. Assume the edge qubits at the $z=0$ plane are in a code state of the toric code $\ket{\phi_\text{TC}}$, and all the other qubits are initialized in the $\ket{+}$ state. First, we apply the unitary $U_\Lambda$ (defined in Eq.~\eqref{eq:ulambda}) -- which we interpret as growing a single layer RBH state -- and then we measure all qubits except the edge qubits on the $z=1$ boundary\footnote{Note that the edge qubits on the bottom boundary are measured as well.} in the Pauli-X basis. Doing so teleports the toric code state from the bottom boundary to the top boundary, i.e.\ the post-measurement state of the top edge qubits (at $z=1$) would be $\ket{\phi_\text{TC}}$, up to $\pm$ sign structure differences in star and plaquette operators which can be easily determined based on the outcomes of the Pauli-X measurements. To see why, first note that one could consider an equivalent alternative scenario, where the edge qubits at $z=0$ start in a toric code logical Bell pair state with a set of added reference qubits, instead of being initialized in $\ket{\phi_\text{TC}}$. Then, the same steps (i.e.\ applying $U_\Lambda$ followed by $X$ measurements) will result in a logical Bell pair between the reference qubits and the edge qubits at the $z=1$ boundary. It is easy to see why this new formulation of the claim is true by viewing the original logical Bell pair between the reference qubits and $z=0$ edge qubits as the post-measurement state of another slab of the RBH state, say between $z=-1$ (where we place the reference qubits) and $z=0$, as described in Appendix~\ref{apx_sub:rbh_Toric_bell}. Then the aforementioned procedure can be seen as making an RBH state on a $L_x\times L_y \times 2$ slab between $z=-1$ and $z=1$ and then measuring all qubits in $X$ basis except those on the top ($z=1$) and bottom ($z=-1$) boundaries. Again, based on Appendix~\ref{apx_sub:rbh_Toric_bell}, this would result in a logical Bell pair between the two boundaries, i.e.\ the reference qubits and the ones on the $z=1$ edges. Then, teleportation follows from measuring logical operators on the reference qubits.

\subsection{The RBH state and repeated syndrome measurements in the absence of noise}

As was explained in the previous section, the RBH state can be used to teleport the toric code along the $z$ axis. It turns out that the procedure outlined above also measures the stabilizers of the toric code. Again, consider the setup with a $L_x\times L_y \times 1$ cubic lattice between $z=0$ and $z=1$ planes, where the edge qubits on the $z=0$ are initially an eigenstate of the toric code plaquette and star stabilizers with a set of unknown eigenvalues $s_i$ (and the rest of the qubits are in the $\ket{+}$ state). Let us first focus on the plaquette operators. Since Pauli-Z commutes with any CZ gate, the resulting state after $U_\Lambda$ is applied remains an eigenstate of the plaquette operators with the same eigenvalue as before. As such, one can read off the value of plaquette stabilizers from the outcome of the Pauli-X measurements performed on the face qubits in the $z=0$ plane (Fig.~\ref{fig:rbh_plaquette}). Next, we consider the star operators. Under the application of $U_\Lambda$, a star operator is dressed by the product of Pauli-Z operators on face qubits that are directly above the edge qubits participating in a star operator, since $\text{CZ}\, XI\, \text{CZ}^\dagger=XZ$. Multiplying the dressed operator by $g_e$, where $e$ is the vertical edge coming out of the star, then shows that the value of the original star operator is now encoded in the five qubit Pauli-X operator shown in Fig.~\ref{fig:rbh_star}. Hence, after measuring qubits in Pauli-X, one can read off the value of the corresponding star stabilizer from the product of the Pauli-X measurement outcomes on these five qubits. 
The above discussion shows that one can describe one round of syndrome measurements in a toric code state as growing a depth-one slab of RBH out of the initial state and then measuring  every qubit, except those on the top edges, in the Pauli-X basis. More generally, $M$ rounds of syndrome measurements can be described in the RBH picture by repeating the same steps $M$ times; the first round of syndrome measurements corresponds to growing a single-layer RBH state followed by Pauli-X measurements except on the top boundary. The next round of syndrome measurements corresponds to growing another layer above the top boundary (between $z=1$ and $z=2$) and then measuring every qubit in Pauli-X except those on the new top boundary, and so on. The edge qubits on the $z=T$ plane then represent the state of the toric code at time $t=T$.  Clearly, one can instead first grow a $L_x\times L_y \times T$ slab of the RBH state and then perform all the Pauli-X measurements throughout the slab. 

\subsection{The RBH state and repeated syndrome measurements in the presence of noise}

\begin{figure}
    \centering
    \includegraphics{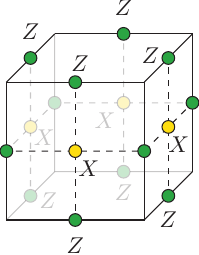} 
    \caption{The product of $g_f$ stabilizers around a cube is equal to the product of toric code plaquette operators on the top and bottom boundary times the product of four $X_f$ in the bulk.}
    \label{fig:rbh_plaquette_teleport}
\end{figure}

In the last section, we explained how one can use a 3D RBH state to represent repeated syndrome measurements of the 2D toric code through time. In this section we explain how the effect of noise (including measurement errors) can be incorporated in the RBH picture. To start, we consider the noise that can happen in one step of toric code syndrome measurements. Therefore we start with a $L_x\times L_y \times 1$ cubic lattice $\Lambda$, where the edge qubits on the $z=0$ plane are initially an eigenstate of plaquette and star operators and rest of the qubits are initialized in the $\ket{+}$ state. As before, we first grow a single-layer RBH state state via by $U_\Lambda$, and then measure all qubits except those on the top boundary in the Pauli-X basis.

First, we explain how plaquette measurement errors can be incorporated into this picture. Since the value of a plaquette operator is inferred from the outcome of the corresponding face qubit at the $z=0$ plane, a phase flip Pauli-Z error on that qubit before the Pauli-X measurements results in a plaquette measurement error. Importantly, such an error does not affect the actual value of the plaquette operator as it is copied from the bottom layer into the top layer after the bulk $X$ measurements. To see this, note that for a given plaquette $p$, the product of $g_f$ where $f$ is a vertical face directly above $p$ is equal to the product of plaquette operators on the top and bottom boundary times the product of $X_f$ on the face qubits in between (Fig.~\ref{fig:rbh_plaquette_teleport}). This stabilizer makes it clear why the value of the bottom plaquette operator is copied to the top layer (up to a sign depending on the $X_f$ measurement outcomes). Since it does not involve the face qubits on horizontal planes, a phase flip on those qubits will not have any effect on the actual value of the plaquette operators on the top boundary after the bulk and bottom boundary qubits are measured in Pauli-X. As for star measurement errors, since their value is inferred from the product of $X$ measurement outcomes shown in Fig.~\ref{fig:rbh_star}, a phase flip Pauli-Z error on the vertical edge qubit directly above the star will result in an erroneous stabilizer measurement. For similar reasons as for the plaquette operators, such an error only flips the stabilizer measurement outcome, leaving unchanged the actual value of the stabilizer on the top boundary. Hence, this correctly represents a star operator measurement error. 

To account for physical errors, it is sufficient to describe how single qubit physical Pauli-X and -Z errors can be represented in the RBH picture. To implement a physical Pauli-Z error on a qubit at time $t=1$, we may simply apply Pauli-Z on the corresponding edge qubit on the top boundary ($z=1$)\footnote{Applying Pauli-Z on the corresponding edge qubit in the $z=0$ plane corresponds to a physical Pauli-Z error at time $t=0$.}. On the other hand, a physical Pauli-X error on a qubit is represented in the RBH picture by a Pauli-Z error on the face qubit directly below the corresponding edge qubit. To see this, recall that the stabilizer shown in Fig.~\ref{fig:rbh_plaquette_teleport} is responsible for copying the value of a plaquette operator from the bottom boundary to the top boundary (once the bulk and bottom boundary qubits have been measured in Pauli-X). A phase flip $Z_f$ on a vertical face qubit flips the sign of this stabilizer and thus flips the sign of the plaquette as it gets copied to the top layer. The same occurs for the adjacent plaquette, as well as the logical $Z$ strings that pass along the corresponding edge qubit (see Fig.~\ref{fig:rbh_zlogical}). Therefore a phase flip $Z_f$ on a face qubit translates into a bit flip $X_e$ on the top edge qubit after the bulk and bottom qubits have been measured in Pauli-X. 

To summarize,
\begin{itemize}
    \item Plaquette stabilizer measurement errors correspond to Pauli-Z errors on horizontal face qubits. 
    \item Star stabilizer measurement errors correspond to Pauli-Z errors on vertical edge qubits. 
    \item Physical Pauli-Z errors correspond to Pauli-Z errors on horizontal edge qubits. 
    \item Physical Pauli-X errors correspond to Pauli-Z errors on vertical face qubits.
\end{itemize}
Importantly, all types of noise in the repeated syndrome measurement setup correspond to Pauli-Z errors in the RBH picture. Since Pauli-Z commutes with CZ gates, in describing $T$ rounds of noisy repeated syndrome measurements one may first grow a $L_x\times L_y \times T$ slab of the RBH state and then, at the end, decohere the bulk according to the noise model. This allows us to remove time completely in the RBH picture, and thus describe the $(2+1)$D dynamics of quantum information via a static $3$D quantum system undergoing one step of decoherence. 

\subsection{Stat-mech model in the stabilizer expansion using the RBH state}
In this section we sketch how one can use the RBH state to arrive at the stat-mech model describing the moments of the density matrix of the 2D toric code under repeated noisy measurement and decoherence in the stabilizer expansion formalism (Sec.~\ref{sec:stat-mech_stabilizer}). The stat-mech model in the error configuration formalism can be obtained via a similar line of reasoning. Moreover, here we focus on the stat-mech model in the bulk and leave out the details of the boundary conditions.

For simplicity, we consider the toric code under only Pauli-X errors with noisy measurement of plaquette operators, which is studied in more detail in Appendix~\ref{app:surface_code}. Moreover, we assume the bit-flip error rate and faulty measurement rate are both equal to a given parameter $p$.  It is straightforward to adapt the following to more general noise channels.

Let $\Lambda^*$ denote the cubic lattice dual to $\Lambda$, where the vertices of $\Lambda$ correspond to cubes of $\Lambda^*$, faces of $\Lambda$ correspond to edges of $\Lambda^*$ and cubes of $\Lambda$ correspond to vertices of $\Lambda^*$. In particular, note that edge stabilizers $g_e$ can be viewed as the face stabilizers for faces in $\Lambda^*$, which we may denote by $g_{f^*}$. Accordingly, the RBH state can be written as,
\begin{align}
    \ketbra{\psi_\text{RBH}}=\sum_{\bm{u},\bm{v}} g^{\bm{u}}g_*^{\bm{v}},
\end{align}
where $\bm{u}$ ($\bm{v}$) is a binary vector specifying a set of faces, i.e.\ membranes, in $\Lambda$ ($\Lambda^*$). $g^{\bm{u}}$ and $g_*^{\bm{v}}$ are shorthand for $\prod_{f\in \Lambda}g_f^{u_f}$ and $\prod_{f^* \in \Lambda^*}g_{f^*}^{v_{f^*}}$ respectively.  In general $g^{\bm{u}}$ ($g_*^{\bm{v}}$) is a Pauli string composed of Pauli-X on qubits on the surface of the membrane specified by $\bm{u}$ ($\bm{v}$) and Pauli-Z on qubits on its boundary. 
According to the discussion above, physical Pauli-X errors in the toric code correspond to Pauli-Z dephasing channels on vertical face qubits of the RBH state and measurement errors on plaquette stabilizers correspond to Pauli-Z dephasing channels on horizontal face qubits of the RBH state. Note that $g_*^{\bm{v}}$ does not have any Pauli-X support on the face qubits. 
Therefore, the corresponding decohered RBH state is,
\begin{align}\label{eq:rbh_dephased}
\calN(\ketbra{\psi_\text{RBH}})=\sum_{\bm{u},\bm{v}} e^{-\mu |\bm{u}|}g^{\bm{u}}g_*^{\bm{v}}
\end{align}
with $\mu=-\log(1-2p)$. Lastly, to faithfully represent the toric code state after multiple rounds of physical noise and noisy measurements, we must measure all the qubits in the bulk of the RBH state projectively in the Pauli-X basis. Note that a projective measurement channel in the Pauli-X basis is the same as a complete Pauli-X dephasing channel. Doing so will restrict the sum in Eq.\eqref{eq:rbh_dephased} to membranes $\bm{u}$ and $\bm{v}$ that do not have any boundaries in the bulk, i.e.\ domain walls, because otherwise $g^{\bm{u}}$ or $g_*^{\bm{v}}$ would have a Pauli-Z support and thus
 be annihilated under the complete Pauli-X dephasing channel. Therefore, we end up with a sum over domain walls in the 3D cubic lattice with energy penalties proportional to the domain wall sizes. This is exactly the 3D Ising model on a 3D cubic lattice, which was derived in Appendix \ref{app:surface_code} as well.

\section{Decoding transition in the repetition code}\label{app:repetition_code}
\subsection{The repetition code}\label{app:repetition_code_intro}
The $d$-dimensional repetition code consists of $N = L^d$ qubits located at the sites of a $d$-dimensional cubic lattice with periodic boundary conditions.
The code has $I = dL^d$ check operators $g_i = Z_{r}Z_{r'}$, which involve qubits on the nearest neighbors $r$ and $r'$ to an edge $i$ in the lattice.
Since $i$ labels the edge between these two sites, we adopt the notation $r,r' \in i$. 
The repetition code encodes one logical qubit of information in the GHZ state over all sites, implying that $I_0 = L^d-1$. 
It follows that the check operators have $I-I_0 = (d-1)L^d+1$ redundancies, each of which is a symmetry of the corresponding stat-mech model. 
The logical $X$ operator is $\prod_r X_r$ and the logical $Z$ operator is $Z_{r_0}$ for any fixed site $r_0$.

\subsection{Stat-mech model}\label{app:repetition_code_model}
The framework developed in Sec.~\ref{sec:stat-mech_stabilizer} enables one to derive the entire $(n-1)$-flavor $(d+1)$-dimensional stat-mech model from just two quantities: $h_r^x(\bm{\sigma}) = \prod_i (\sigma_i)^{a_{i,r}^x}$ and $h_r^z(\bm{\sigma}) = \prod_i (\sigma_i)^{a_{i,r}^z}$. These terms couple together all the check operators with Pauli-X and Pauli-Z support, respectively, at site $r$. It follows that $h_r^x = 1$ and $h_r^z = \prod_{i \ni r} \sigma_i$. In other words, no check operators have Pauli-X support at any site and all check operators at edges $i$ such that $r \in i$ have Pauli-Z support at $r$.

Since the repetition code only protects quantum information against Pauli-X errors, we consider the case where $p_y = p_z = 0$, which implies $J_y = J_z = 0$. Then the Hamiltonian for each flavor is
\begin{equation}\label{eq:repetition_code}
    \calH(\bm{\sigma}) = -J_x\sum_{t,r} \prod_{i \ni r} \sigma_{t,i} - J_q\sum_{t,i} \sigma_{t,i}\sigma_{t+1,i},
\end{equation}
where $J_q = -(1/2)\log(1-2q)$.
The Hamiltonian $\calH^{(n)}(\{\bm{\sigma}^\sfa\})$ follows from this, as defined in Eq.~\eqref{eq:stabilizer_expansion_flavor_hamiltonian}. For the 1D repetition code, the corresponding stat-mech model is a 2D Ising model. For the 2D repetition code, it is a 3D $\bbZ_2$ gauge theory.\footnote{In each layer, spins on edges which share a vertex are coupled, which looks like a four-body plaquette coupling on the dual lattice. Then, Eq.~\eqref{eq:repetition_code} describes layers of 2D $\bbZ_2$ gauge theories with Ising couplings between adjacent layers. This is equivalent to a 3D $\bbZ_2$ gauge theory with all temporal links set to $+1$, which is always possible using local gauge transformations.} Neither model orders at finite temperature when $J_q = 0$ (in which case they are reduced to a 1D Ising model and a 2D $\bbZ_2$ gauge theory respectively), indicating that they have maximum thresholds, i.e.\ $p_c = 0.5$, in the absence of readout noise. In the presence of readout noise, both models have sub-maximum thresholds, i.e.\ $p_c < 0.5$.

The symmetry group resulting from check operator redundancies is $\calR \cong \bbZ_2^{I-I_0}$ for each flavor and $(\calR^n\rtimes S_n)/\calR$ for the full stat-mech model. As noted above, $I-I_0 = (d-1)L^d+1$ for the $d$-dimensional repetition code. When $d=1$, this corresponds to the global $\bbZ_2$ symmetry in each flavor of the Ising model. For $d>1$ there are an extensive number of local symmetries generated by the spin flips of the spins along the edges of the plaquettes in the original lattice.

\subsection{Information diagnostics}
Here, we show the transition in the stat-mech model $\calZ_n$ manifests in the information diagnostics.
First, the coherent information depends on the stat-mech model we determined in Appendix~\ref{app:repetition_code_model} as well as related partition functions with defects inserted. 
In particular, for $\bm{\kappa} = (\kappa_x,\kappa_z) \in \bbZ_2^2$ with $\kappa_x$ and $\kappa_z$ indicating the insertion of the logical $X$ and $Z$ operator, the Hamiltonian terms $h_{\bm{\kappa},r}^x$ and $h_{\bm{\kappa},r}^z$ change accordingly. 
We find that $h_{\bm{\kappa},r}^{x} = (-1)^{\kappa_x}$ since the logical $X$ operator has Pauli-X support on every site, and $h_{\bm{\kappa},r}^z = (-1)^{\kappa_z}\delta_{r,r_0}\prod_{i \ni r} \sigma_i$ since the logical $Z$ operator has Pauli-Z support only on site $r_0$. 
Here, only $h_{\bm{\kappa},r}^z$ appears in the Hamiltonian when $p_y = p_z = 0$, in which case we have 
\begin{equation}
    \calH_{\bm{\kappa}}(\bm{\sigma}) =  -J_x\sum_{t,r} (-1)^{\kappa_z}\delta_{r,r_0}\prod_{i \ni r} \sigma_{t,i} - J_q\sum_{t,r} \sigma_{t,i}\sigma_{t+1,i}.
\end{equation}
The logical sector $\kappa_z = 1$ corresponds to inserting a temporal defect at site $r_0$ along which $\mu_x \to -\mu_x$. 

The R\'enyi-$n$ coherent information then takes the form
\begin{equation}
    I_c^{(n)} = \frac{1}{n-1}\log(\sum_{\{\kappa_z^\sfa\}} e^{-\Delta F_n(\{\kappa_z^\sfa\})}),
\end{equation}
where the summation over $\kappa_z^\sfa$ runs over $2^{n-1}$ choices of inserting temporal defects at site $r_0$ in each flavor.
Here, we have performed the summation over $\kappa_x^\sfa$, which cancels the constant term in Eq.~\eqref{eq:renyi-n_coherent_info_stat-mech}.
The R\'enyi coherent information can then detect the transition in the stat-mech model; $I_c^{(n)} = \log 2$ when each defect insertion costs vanishing free energy, i.e.\ $\Delta F_n(\{\kappa_z^\sfa\}) \to 0$, and $I_c^{(n)} = 0$ when each defect has a cost $\Delta F_n(\{\kappa_z^\sfa\}) \to \infty$. 
For the 1D repetition code, $\Delta F_n(\{\kappa_z^\sfa\}) \to 0$ in the paramagnetic phase of the corresponding 2D Ising model and $\Delta F_n(\{\kappa_z^\sfa\}) \to \infty$ in the ferromagnetic phase. 
For the 2D repetition code, $\Delta F_n(\{\kappa_z^\sfa\})$ is the excess free energy for inserting a magnetic $\pi$-flux line at position $r_0$ in the temporal direction.
In the corresponding 3D $\bbZ_2$ gauge theory, the excess free energy vanishes in the deconfined phase, while it is of $O(T)$ and diverges in the confined phase.

The phase transition in the stat-mech model is also probed by the relative entropy between $\rho_{Q\bfM}$ and $\rho_{Q\bfM,\bm{s}}$ where $\bm{s}$ denotes the syndromes created by applying Pauli $X_r$ to a contiguous region in the system.
For the 1D repetition code, we consider applying Pauli-X to a subregion such that $s_i = s_j = -1$ for $i$ and $j$ on the boundary. 
Then, the R\'enyi relative entropy is given by the two-point correlation function at the bottom of the stat-mech model, i.e.\ $D_{\bm{s}}^{(n)} = -\frac{1}{n-1}\log\expval{\sigma^1_{0,i}\sigma^1_{0,j}}$.
It grows with the size of the subregion $|i-j|$ in the paramagnetic phase, while it remains constant in the ferromagnetic phase. 
For the 2D repetition code, we consider applying Pauli-X to a subregion $\mathcal{A}$ generating $s_i = -1$ for all $i$ along a loop $\partial \mathcal{A}$ (on the dual lattice along the boundary of the subregion $\mathcal{A}$). 
Then, $D_{\bm{s}}^{(n)} = -\frac{1}{n-1}\log\expval{\prod_{i\in \partial \mathcal{A}}\sigma^1_{0,i}} = -\frac{1}{n-1}\log\langle e^{\ri\oint_{\partial \calA} A \rd \ell}\rangle$ which grows with the area of $\calA$ in the area-law (confined) phase and with $|\partial \calA|$ in the perimeter-law (deconfined) phase.

\section{Decoding transition in XZZX code}\label{app:XZZX}
\subsection{The XZZX code}\label{app:XZZX_intro}
\begin{figure}[t!]
    \centering
    \begin{tikzpicture}[]
    \pgftext{\includegraphics[width=0.49\linewidth]{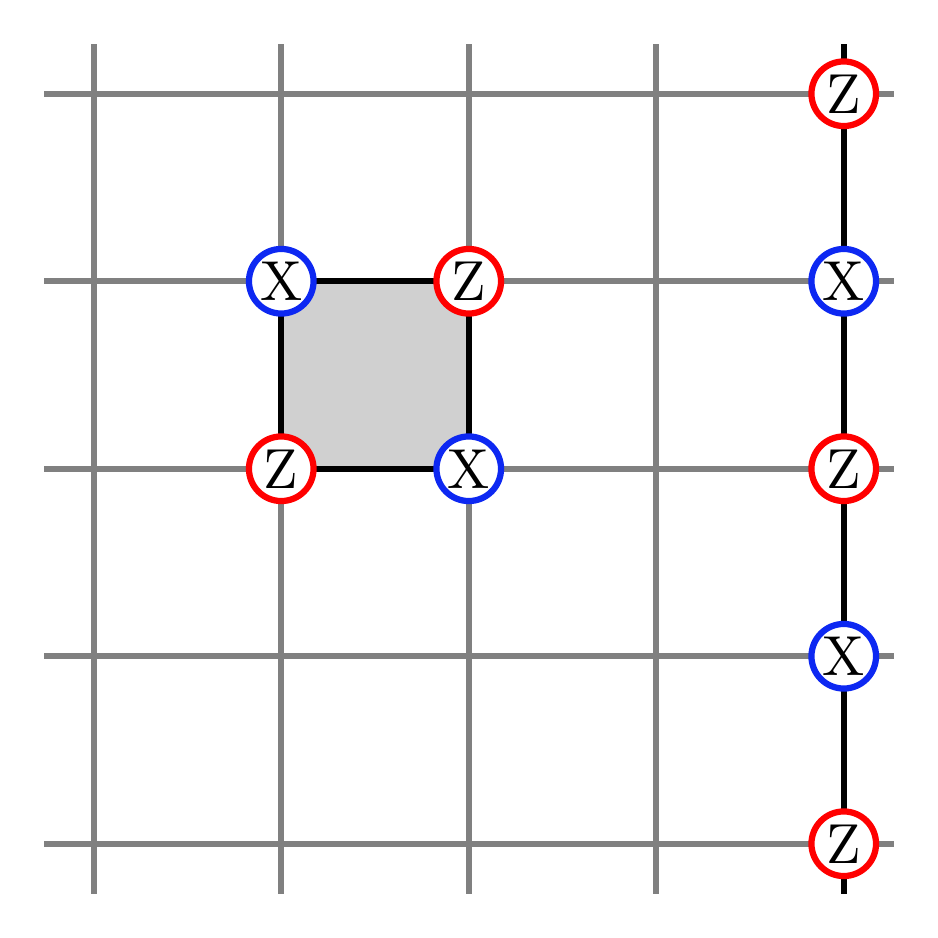}} at (0pt,0pt);
    \node at (-1.3, 0.4) {(a)};
    \node at (1.25, 0.4) {(b)};
    \end{tikzpicture}
    \caption{The XZZX code. (a)~Each stabilizer is a four-body Pauli operator associated with a plaquette. Each has Pauli-X operators on the top-left and bottom-right corners and Pauli-Z operators on the top-right and the bottom-left corners. (b)~The logical operators are Pauli strings along non-contractible loops involving products of Pauli-X and -Z operators. }
    \label{fig:XZZX_intro}
\end{figure}

The XZZX code \cite{bonilla2021xzzx} involves qubits on the vertices of the square lattice, which is equivalent to a toric code up to local rotations at certain sites. It is notable for saturating the maximum error threshold of $p_c=0.5$ when subject to only one of Pauli-X, -Y, or -Z physical errors.
The XZZX code has a stabilizer group generated by the products of four Pauli operators on the vertices of a plaquette and encodes two logical qubits.
Examples of stabilizer and logical operators are illustrated in Fig.~\ref{fig:XZZX_intro}.

Here, we consider the XZZX code on a $L \times L$ square lattice with periodic boundary conditions and discuss the stat-mech model for the code subject to Pauli-X, -Y, or -Z errors and repeated noisy syndrome measurements on all plaquettes.
The set of check operators involving all plaquettes exhibits an $\calR = \bbZ_2\times \bbZ_2$ redundancy, giving rise to the symmetry of the stat-mech model.

\subsection{Pauli-X and -Z errors}
 
In this section, we determine $\calZ_n$ for the XZZX code subject to Pauli-X or -Z errors.  
As shown in Sec.~\ref{sec:stat-mech_stabilizer}, the only code-specific aspects of $\calZ_n$ are the local spin terms $h_r^x$ and $h_r^z$, so it is sufficient to compute only these terms.

\begin{figure}[t!]
    \centering
    \begin{tikzpicture}[]
    \pgftext{\includegraphics[width=0.49\linewidth]{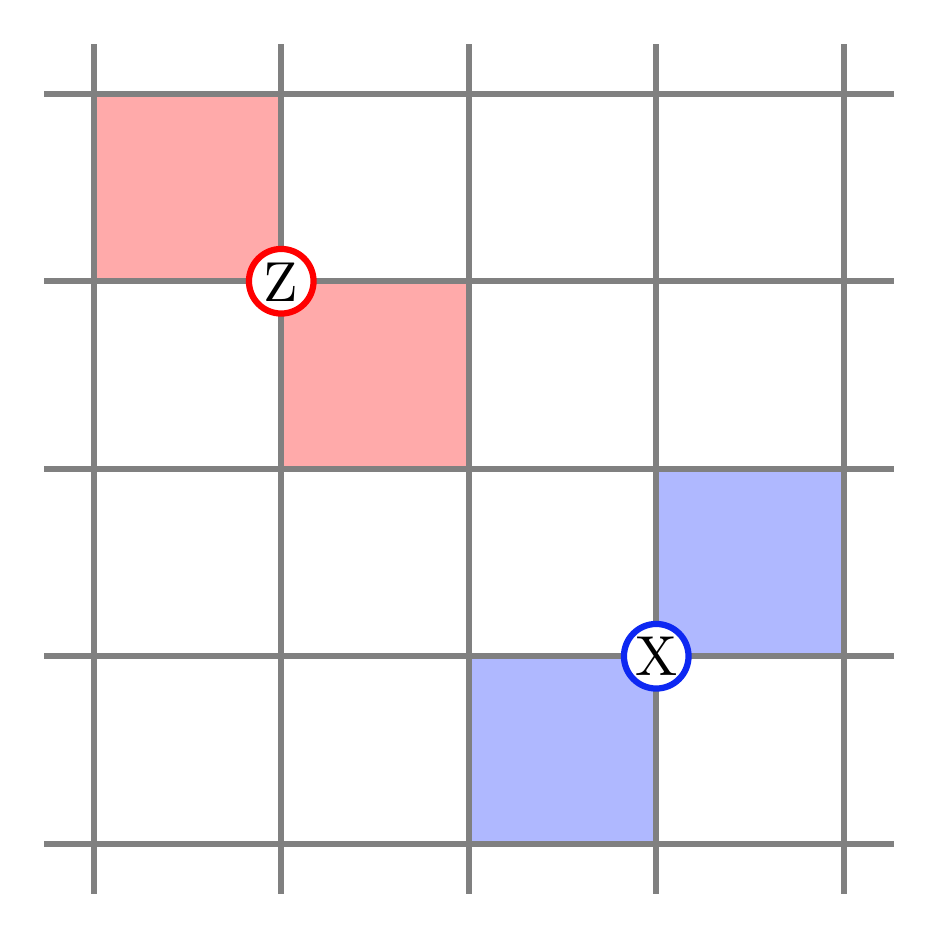}\includegraphics[width=0.49\linewidth]{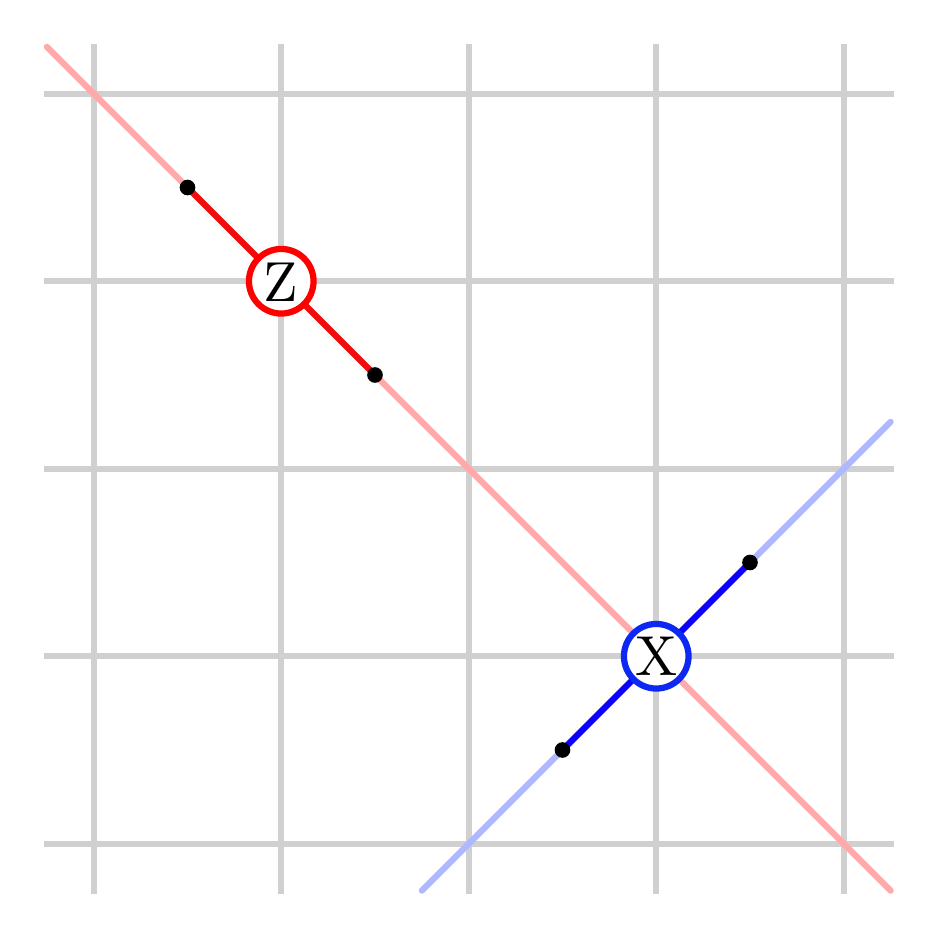}} at (0pt,0pt);
    \node at (-2, -2.2) {(a)};
    \node at (2, -2.2) {(b)};
    \end{tikzpicture}
    \caption{(a)~Pauli-Z error at a site anticommutes with two diagonally-adjacent plaquette operators (colored red) that have Pauli-X support at that site. Corresponding plaquettes along the opposite diagonal have Pauli-Z support at a site, leading them to anticommute with Pauli-X errors. (b)~Pauli-Z error induces two-body Ising couplings $h_r^x$ along the diagonal from the top left to the bottom right. The analogous statement is true for Pauli-X error and $h_r^z$, with plaquettes coupled along the opposite diagonal.}
    \label{fig:XZZX_X,Z}
\end{figure}

As illustrated in Fig.~\ref{fig:XZZX_X,Z}(a), the upper-left and lower-right plaquettes at a site have Pauli-X support at that site, which anticommutes with Pauli-Z errors. Likewise, the upper-right and lower-left plaquettes have Pauli-Z support at that site, which anticommutes with Pauli-X errors. Taking the bottom left to be $x=y=0$, we conclude that $h_r^x(\bm{\sigma}) = \sigma_{(x,y)}\sigma_{(x+1,y-1)}$ and $h_r^z(\bm{\sigma}) = \sigma_{(x,y)}\sigma_{(x+1,y+1)}$. 

Therefore, when only Pauli-Z errors are present, each flavor of the stat-mech model consists of $L$ uncoupled 1D Ising models running along diagonal lines from the top-left to the bottom-right. Similarly, when only Pauli-X errors are present, each flavor of the stat-mech model consists of $L$ uncoupled 1D Ising models running along diagonal lines from the top-right to the bottom-left. These systems only order at zero temperature, leading to a decoding threshold at $p_{z,c}=0.5$ (or $p_{x,c} = 0.5$) when there is no readout noise. Representative couplings and diagonal lines are illustrated in Fig.~\ref{fig:XZZX_X,Z}(b). 
In the presence of readout noise, the stat-mech model instead consists of decoupled 2D Ising models and exhibits a finite-temperature phase transition, giving rise to a sub-maximum error threshold, i.e.\ $p_c < 0.5$.

As derived in Sec.~\ref{sec:stat-mech_stabilizer} and demonstrated in Appendix~\ref{app:surface_code} and Appendix~\ref{app:repetition_code}, the coherent information may be formulated explicitly in terms of free energy costs of defect insertions. These defect insertions arise from the Pauli-X and Pauli-Z support of the logical operators. In particular, each logical operator flips the sign of the coupling on one bond in each 1D Ising model.

\subsection{Pauli-Y errors}
\begin{figure}[t!]
    \centering
    \begin{tikzpicture}[]
    \pgftext{\includegraphics[width=0.49\linewidth]{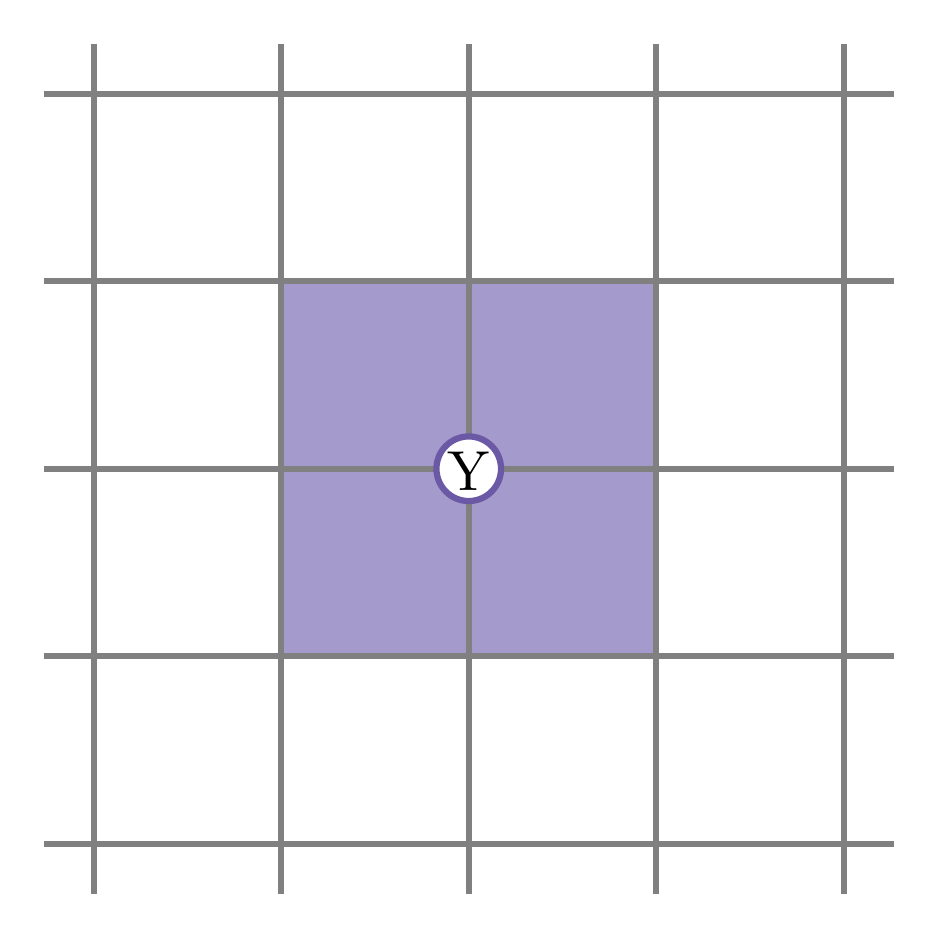}\includegraphics[width=0.49\linewidth]{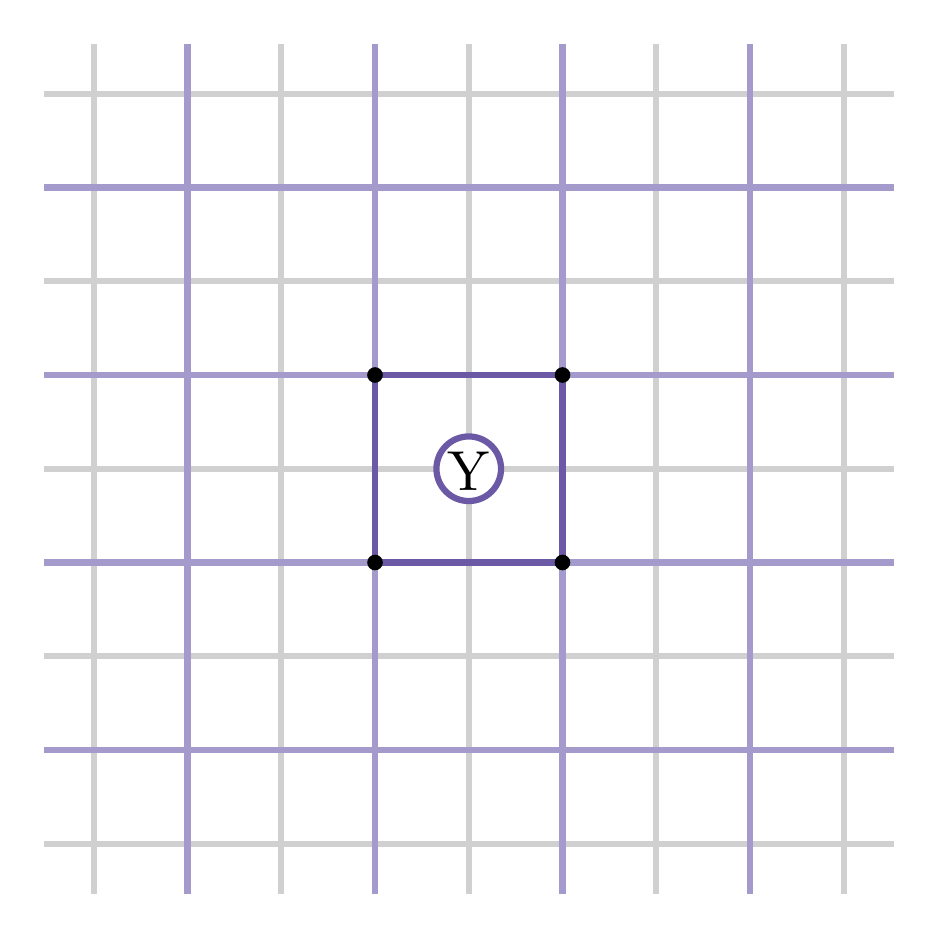}} at (0pt,0pt);
    \node at (-2, -2.2) {(a)};
    \node at (2, -2.2) {(b)};
    \end{tikzpicture}
    \caption{(a)~Pauli-Y error at a site anticommutes with all four adjacent plaquettes, each of which has Pauli-X or Pauli-Z support at that site. (b)~The result is a four-body plaquette Ising coupling on the lattice of classical spins.}
    \label{fig:XZZX_Y}
\end{figure}

When Pauli-Y errors are present, a four-body coupling $h_r^xh_r^z = \prod_{i \in \Box} \sigma_i$ is added to the Hamiltonian, where $\Box$ denotes the vertices of the plaquette dual to $r$. The implicated check operators and the resulting coupling are illustrated in Fig.~\ref{fig:XZZX_Y}(a) and Fig.~\ref{fig:XZZX_Y}(b), respectively.

When only Pauli-Y error is present and there is no readout noise, the decodability of the XZZX code is thus captured by a $2$D plaquette Ising model, like the surface code under Pauli-Y error. This model does not order at finite temperature, giving rise to a maximum decoding threshold of $p_{y,c}=0.5$, as expected.
When we include readout error, the model is $T$ layers of plaquette Ising models coupled by two-body Ising interactions. 
This model generally orders at a finite temperature, leading to a sub-maximum error threshold.

\end{document}